\documentclass[11pt,a4paper]{article}
\usepackage{jcappub}

\usepackage{array,calc,epsfig}
\usepackage{bbm}
\usepackage{fancybox}
\usepackage{xcolor}

\newcommand{\md}{\mathrm{d}}
\def\be{\begin{equation}}
\def\ee{\end{equation}}
\def\bseq{\begin{subequations}}
\def\eseq{\end{subequations}}

\def\bea{\begin{eqnarray}}
\def\eea{\end{eqnarray}}

\def\bseq{\begin{subequations}}
\def\eseq{\end{subequations}}

\title{Towards a viable effective field theory of mimetic gravity}
\author[a,b]{Alexander Ganz,}
\author[a,b,c]{Nicola Bartolo,}
\author[a,b,c,d]{Sabino Matarrese}
\affiliation[a]{ Dipartimento di Fisica e Astronomia ``G. Galilei", \\
	Universit\`a degli Studi di Padova, via Marzolo 8, I-35131 Padova, Italy}
\affiliation[b]{INFN, Sezione di Padova, \\ via Marzolo 8, I-35131 Padova, Italy}
\affiliation[c]{INAF - Osservatorio Astronomico di Padova, \\ Vicolo dell’Osservatorio 5, I-35122 Padova, Italy}
\affiliation[d]{Gran Sasso Science Institute, \\ Viale F. Crispi 7, I-67100 L'Aquila, Italy}
\emailAdd{alexander.ganz@pd.infn.it}
\emailAdd{nicola.bartolo@pd.infn.it}
\emailAdd{sabino.matarrese@pd.infn.it} 

\abstract{We discuss mimetic gravity theories with direct couplings between the curvature and higher derivatives of the scalar field, up to the quintic order, which were proposed to solve the instability problem for linear perturbations around the FLRW background for this kind of models. Restricting to homogeneous scalar field configurations in the action, we derive degeneracy conditions to obtain an effective field theory with three degrees of freedom. However, performing the Hamiltonian analysis for a generic scalar field we show that there are in general four or more degrees of freedom. The discrepancy is resolved because, for a homogeneous scalar field profile, $\partial_i\varphi\approx 0$, the Dirac matrix becomes singular, resulting in further constraints, which reduces the number of degrees of freedom to three. Similarly, in linear perturbation theory the additional scalar degree of freedom can only be seen by considering a non-homogeneous background profile of the scalar field. Therefore, restricting to homogeneous scalar fields these kinds of models provide viable explicitly Lorentz violating effective field theories of mimetic gravity. 
}

\arxivnumber{}

\notoc 

\begin{document}
\maketitle
\flushbottom
\section{Introduction}
In \cite{Chamseddine:2013kea} a modification of General Relativity (GR) was proposed by performing a non-invertible conformal transformation of the Einstein-Hilbert action
\begin{align}
    g_{\mu\nu} = - \left(\tilde g^{\alpha\beta} \partial_\alpha \varphi \partial_\beta \varphi \right) \tilde g_{\mu\nu}.
\end{align}
The additional scalar degree of freedom of the conformally invariant theory, called ``mimetic gravity", can mimic the behavior of a pressureless fluid. Later, it was realized \cite{Barvinsky:2013mea,Golovnev:2013jxa} that by gauge fixing the conformal symmetry, we can obtain mimetic gravity by adding the mimetic constraint
\begin{align}
    \label{eq:Mimetic_constraint_introduction}
    g^{\mu\nu} \partial_\mu\varphi \partial_\nu\varphi +1 =0
\end{align}
via a Lagrange multiplier term 
to the action. Mimetic gravity can be generalized by adding the mimetic constraint to a generic scalar-tensor theory \cite{Chamseddine:2014vna,Arroja:2015wpa,Arroja:2015yvd,Arroja:2017msd,Takahashi:2017pje,Langlois:2018jdg} (see \cite{Sebastiani:2016ras} for a review).

 In \cite{Langlois:2018jdg,Takahashi:2017pje,Zheng:2017qfs,Firouzjahi:2017txv,Ijjas:2016pad,Hirano:2017zox,Gorji:2017cai,Ganz:2018mqi} linear perturbations around a Friedmann-Lema\^itre-Robertson-Walker (FLRW) background (FLRW) for a wide range of mimetic gravity models were analyzed. It was argued that these theories might be plagued with gradient or ghost instabilities, which can be interpreted in the original mimetic matter model as the Jeans instability of standard dust \cite{Ganz:2018mqi}. To circumvent the instability problems it was proposed to add direct 
 couplings of higher-order derivative terms of the scalar field to the curvature \cite{Hirano:2017zox,Zheng:2017qfs,Gorji:2017cai}.
 
 By performing the Hamiltonian analysis it was shown that the original mimetic matter model has three degrees of freedom (dof), due to the presence of the conformal symmetry, or the mimetic constraint in the gauge-fixed version \cite{Chaichian:2014qba,Malaeb:2014vua}. The analysis was then generalized for a broad range of mimetic gravity theories, starting with general scalar-tensor theories containing second-order derivatives of the scalar field \cite{Takahashi:2017pje,Zheng:2018cuc,Ganz:2018mqi}. However, it was argued \cite{Takahashi:2017pje} that a direct coupling of 
 higher-order derivative terms of the scalar field to the curvature may generate an additional scalar degree of freedom (dof) which will cause an Ostrogradski ghost instability \cite{Woodard:2015zca}. In \cite{Zheng:2018cuc} a mimetic gravity theory containing $f(\Box \varphi) R$ was analyzed showing that there will be in general four degrees of freedom. 
 However, if one identifies the scalar field with the time flow $\varphi=t$, which is commonly called ``unitary gauge", the additional dof vanishes leaving just three dof.
 
 A similar situation, where the choice of the form of the scalar field can change the number of dof has been encountered in a series of modified gravity models, such as Horava-Lifshitz, Cuscuton models, higher-order scalar-tensor theories (HOST), or modified Chern-Simons gravity models \cite{Blas:2009yd,Chagoya:2018yna,Gomes:2017tzd,DeFelice:2018mkq,Iyonaga:2018vnu,Crisostomi:2017ugk}. 
 In \cite{DeFelice:2018mkq} it was argued that for degenerate HOST models the additional non-propagating ghost degree of freedom in a non-unitary gauge can be removed by imposing the regularity condition at spatial infinity, while in the unitary gauge this boundary condition is already imposed implicitly by the choice of the gauge itself. 
 In \cite{Gomes:2017tzd} the Hamiltonian for the Cuscuton was studied, showing that, for generic field configurations, there are three dof which reduce to two by imposing the unitary gauge strongly at the Lagrangian level. It was argued that the homogeneous limit is singular and discontinuous from the general case and, therefore, both cases have to be treated separately as two different dynamical systems. 
 
 In this paper we address some of these points with the goal of revisiting the effect of the homogeneity condition of the scalar field on the number of degrees of freedom for mimetic gravity theories, with couplings of higher-order derivatives of the scalar field to the curvature, hence providing possible viable effective field theories of mimetic gravity. In section \ref{sec:Higher_Derivatives} we analyze the higher-derivative operators proposed in \cite{Gorji:2017cai}, for homogeneous field configurations, $\varphi=t$, and derive the degeneracy conditions to obtain a theory with three dof.
 In section \ref{sec:Hamiltonian_analysis} we perform the Hamiltonian analysis for generic scalar fields for the cubic operators (sketching higher orders in the appendix), showing that there will be in general four dof. Imposing the homogeneity condition, $\varphi\approx t$, the Dirac matrix becomes singular leading to two further second-class constraints which reduce the number of dof by one, thereby generalizing the results
 of \cite{Zheng:2018cuc} to a broad range of operators.
 In the last section, we study, for a particular cubic operator, linear scalar perturbations around flat space, both for homogeneous and inhomogeneous background scalar field configurations. We show that in the case of the non-homogeneous background scalar field there are two scalar degrees of freedom from which at least one of them is a ghost. 
 
 Some tedious calculations and additional comments are given in the appendix. We take the $(-,+,+,+)$ signature and choose units where the speed of light and the reduced Planck mass are set to one. 
 Greek indices run from 0 to 3 and Latin indices from 1 to 3.
 
\section{Higher-derivative coupling operators}
\label{sec:Higher_Derivatives}
By introducing higher-derivative operators coupled to the curvature one can obtain stable perturbations around FLRW for mimetic gravity scenarios \cite{Zheng:2017qfs,Hirano:2017zox,Gorji:2017cai}. Here, we want to discuss these operators and especially the associated number of the degrees of freedom, in the case in which the mimetic scalar field is identified with the time flow $\varphi=t$. 

\subsection{Relation between curvature tensors}
\label{subsec:Relation_curvature}
We shortly review the relations between the 4-dimensional (4-dim) Riemann $R_{\alpha\beta\mu\nu}$ tensor, its 3-dim counterpart $\bar R_{\alpha\beta\mu\nu}$ on the hypersurface of constant time and the extrinsic curvature $K_{\mu\nu}$. For the foliation of spacetime we use the standard convention
\begin{align}
    g_{\mu\nu} = h_{\mu\nu} - n_\mu n_\nu,
\end{align}
where $h_{\mu\nu}$ is the metric on the hypersurface of constant time and $n_\mu$ is its normal vector. Further, we use the sign convention
\begin{align}
    \nabla_\mu n_\nu = K_{\mu\nu} - n_\mu a_\nu,
\end{align}
where $a_\nu$ is the acceleration vector and $K_{\mu\nu}$ is the extrinsic curvature. The Gauss, Codazzi  and Ricci relations of the curvature are given by
\begin{align}
    h_\mu^\epsilon h_\delta^\nu h_\rho^\alpha h_\sigma^\beta R^\mu_{\phantom\mu \nu\alpha\beta}&= \bar R^\epsilon_{\phantom\epsilon \delta\rho\sigma} + K^\epsilon_\rho K_{\delta\sigma} - K^\epsilon_\sigma K_{\rho\delta}, \\
     h_\mu^\epsilon h_\delta^\nu h_\rho^\alpha  R_{\epsilon\nu\alpha\beta} n^\beta &= D_\mu K_{\delta \rho} - D_\delta K_{\mu\rho}, \\
     R_{\alpha\beta\mu\nu} n^\beta n^\nu &= - \mathcal{L}_n K_{\alpha\mu} + K_{\alpha\sigma} K^{\sigma}_\mu + D_{(\alpha} a_{\mu)} +a_\alpha a_\mu,
\end{align}
where $\mathcal{L}_n$ is the Lie-derivative along $n^\mu$, $D_\alpha$ is the covariant derivative induced by $h_{\alpha\beta}$ and $2 D_{(\alpha} a_{\beta)}= D_\alpha a_\beta + D_\beta a_\alpha\,$. By contracting the above equations, we obtain relations for the Ricci tensor and Ricci scalar:
\begin{align}
\label{eq:Gauss_equation}
    h^\mu_\alpha h^\nu_\beta R_{\mu\nu} =& \bar R_{\alpha\beta} + \mathcal{L}_n K_{\alpha\beta} + K K_{\alpha\beta} - 2 K_{\alpha\sigma} K^\sigma_\beta   - D_{(\alpha} a_{\beta)} - a_\alpha a_\beta, \\
    R_{\mu\nu} h^\mu_\alpha n^\nu =& D_\mu K_\alpha^\mu - D_\alpha K, \\
    R_{\mu\nu} n^\mu n^\nu =& K^2 - K_{\alpha\beta } K^{\alpha\beta} + \nabla_\alpha \left( a^\alpha - n^\alpha K \right), \\
    R =&  \bar R + K_{\alpha\beta} K^{\alpha\beta} - K^2 -2 \nabla_\alpha  \left( a^\alpha - n^\alpha K \right).
\end{align}

In the following we will parametrize the metric by the specific choice of the ADM decomposition \cite{Arnowitt:1962hi}
\begin{equation}
\mathrm{d} s^2 = - N^2 \md t^2 + h_{ij} \left( \md x^i + N^i \md t\right) \left( \md x^j + N^j \md t\right),
\end{equation}
where $N$ is the lapse and $N^i$ the shift vector. For this coordinate choice the extrinsic curvature and the acceleration vector are given by
\begin{align}
    K_{ij} = \frac{1}{2 N} \left( \dot h_{ij} - 2 D_{(i} N_{j)} \right), \qquad a_i = D_i \log N.
\end{align}

\subsection{Unitary gauge}
Identifying the time flow with the scalar field $\varphi=t$, we can define the normal vector $n_\mu$ with respect to the hypersurface of constant time as
\begin{align}
    n_\mu = \frac{- \nabla_\mu \varphi}{\sqrt{-\nabla_\nu \varphi \nabla^\nu\varphi}}= - N \delta_\mu^0.
\end{align}
Further, the mimetic constraint \eqref{eq:Mimetic_constraint_introduction} reduces to
\begin{align}
   g^{\mu\nu} \partial_\mu\varphi \partial_\nu \varphi +1=  - \frac{1}{N^2} + 1=0 \, ,
\end{align}
fixing therefore the lapse $N^2=1$. For simplicity, in the following we will take the positive sign for the lapse function. For a generic action 
\begin{align}
\label{eq:Unitary_Gauge_generic_action}
    S = \int \md^4x\;  \left[ N \sqrt{h} \mathcal{L} - \sqrt{h} \lambda \left( - \frac{1}{N} +  N \right) \right]
\end{align}
the equation of motion (EOM) for the lapse $N$ fixes the Lagrange parameter $\lambda$ 
\begin{align}
    2 \lambda= - \sum_i (-1)^{i+1} \frac{\mathrm{d}^i}{\mathrm{d} t^i}\frac{\delta \mathcal{L}}{\delta \partial_t^{i} N}\Big\vert_{N=1} +  \frac{\delta \mathcal{L}}{\delta N}\Big\vert_{N=1} + \sum_i (-1)^{i} \frac{\partial^i}{(\partial x^j)^i} \frac{\delta \mathcal{L}}{\delta (\partial x^j)^i N} \Big\vert_{N=1} .
\end{align}
Note, that any term in the Lagrangian, which is quadratic in derivatives of the lapse, vanishes in the EOM, after imposing the mimetic constraint. 

We can also directly plug the mimetic constraint $N=1$ into the action \cite{Hirano:2017zox} and obtain
\begin{align}
\label{eq:Mim_action_Unitary_Gauge}
    S = \int \md^4x\,  \sqrt{h} \mathcal{L} \vert_{N=1}.
\end{align}
Or, alternatively, starting from the generic action \eqref{eq:Unitary_Gauge_generic_action} we can expand the Lagrangian in terms of $N-1$ as a Taylor series. As a next step, we can redefine the Lagrange parameter $\lambda$ to absorb all the terms proportional to $N-1$ (see \cite{Ganz:2018mqi,Ganz:2018vzg} for a similar discussion for the general mimetic constraint). The action can be finally expressed as
\begin{align}
    S = \int \md^4x\, \sqrt{h} \Big[ \mathcal{L}\vert_{N=1} - \tilde \lambda \left( N - 1 \right) \Big],
\end{align}
which is classically equivalent to \eqref{eq:Mim_action_Unitary_Gauge}. 

\subsection{Cubic operators}
\label{subsec:Cubic_operators}
Considering the coupling between the higher-order derivatives of the scalar field and the curvature we analyze, at first, cubic operators $L^{(1,2)}$, whereby the classification is based on the second derivative of the scalar field (i.e. $\Box\varphi$ is a linear operator and $(\Box\varphi)^2$ a quadratic one). Further, the curvature terms as $R$, $R_{\mu\nu}$ or $R_{\mu\nu\alpha\beta}$ are identified as quadratic terms. The index $(1,2)$ denotes the order of the operator, where the first index comes from the derivatives of the scalar field and the second one from the curvature (see \cite{Gorji:2017cai} for a detailed discussion).

There are only three independent cubic operators which contribute 
to the EOM of mimetic gravity theories
\begin{align}
     L_1^{(1,2)}= a_1(\varphi) \varphi^{\mu\nu} R_{\mu\nu}, \quad L_2^{(1,2)}=a_2(\varphi) \Box\varphi R, \quad L_3^{(1,2)}= a_3(\varphi) \Box\varphi \varphi^\mu \varphi^\nu R_{\mu\nu}.
\end{align}
Other operators, as mentioned in \cite{Gorji:2017cai}, do not contribute to the EOM, since they contain covariant derivatives of $X$ and vanish identically by using the mimetic constraint (see \cite{Langlois:2018jdg,Ganz:2018mqi} for a detailed discussion).

As discussed in \cite{Langlois:2018jdg}, $L_1^{(1,2)}$ can be recast into
\begin{align}
    L_1^{(1,2)}= - \frac{a_{1}^\prime}{2} R + a_{1}^\prime \left( \varphi_{\mu\nu} \varphi^{\mu\nu} - \left( \Box\varphi \right)^2 \right) + a_{1}^{\prime\prime} \Box\varphi + \nabla_\mu J^\mu,
\end{align}
where $a_1^\prime \equiv \partial_\varphi a_1$ and $\nabla_\mu J^\mu$ stands for a total derivative term. Since all these terms are already commonly discussed for mimetic gravity scenarios, they provide no new information 
and, in particular, they cannot solve the instability problem around the FLRW background. Therefore, we will not consider $L_1^{(1,2)}$ further.

In the previous section we showed that for the choice of $\varphi=t$ the mimetic constraint fixes the lapse function $N=1$ and eventually we are only interested in the restricted operators with $L_i^{(1,2)}\vert_{N=1}$. Up to total derivatives these operators can then be rewritten as
\begin{align}
    L_2^{(1,2)}\vert_{N=1}& = - a_2 E ( \bar R + E_{ij} E^{ij} - E^2 ) -2 a_2 E \nabla_\alpha ( n^\alpha E) \nonumber \\
    &= - a_2 E ( \bar R + E_{ij} E^{ij} ) + a_{2}^\prime E^2 , \\
    L_3^{(1,2)} \vert_{N=1} &  = a_3 E \left( E_{ij} E^{ij} - \frac{1}{2} E^2 \right) - \frac{1}{2} a_3^\prime E^2 ,
\end{align}
where we have introduced the convention $E_{ij}\equiv N K_{ij}=( \dot h_{ij} - 2 D_{(i} N_{j)} )/2$. 

We can observe that by choosing the hypersurface of constant time as $\varphi=t$ both operators can be expressed as purely spatial covariant combinations of the extrinsic curvature and the 3-dimensional Ricci scalar. Performing the Hamiltonian analysis we obtain six first-class constraints, corresponding to time-independent spatial transformations. Having 18 canonical conjugate variables $h_{ij}$, $\pi_{ij}$, $N_j$ and $\pi_j$ (see subsection \ref{subsec:Hamiltonian_analysis_cubic_order} for the chosen convention) we obtain three degrees of freedom. These kinds of operators are specific sub-classes of the spatial covariant gravity theories \cite{Gao:2018znj,Gao:2014fra,Saitou:2016lvb} with the additional condition on the lapse function $N=1$. The condition on the lapse function is reminiscent of the projectable Horava Lifshitz gravity theories \cite{Horava:2009uw} (see also \cite{Ramazanov:2016xhp}, for the connection between the infrared limit of projectable Horava Lifshitz gravity and mimetic gravity).

Finally, note that the relevant part of the term $L_3^{(1,2)}$ for $\varphi=t$ can be equivalently obtained from 
\begin{align}
  \left[ a_3 \left(   \frac{1}{2} (\Box\varphi)^3 - \varphi_{\alpha\beta} \varphi^{\alpha\beta} \Box \varphi \right) - \frac{1}{2} a_3^\prime (\Box\varphi)^2 \right] \Big\vert_{N=1} = a_3  E \left( E_{\alpha\beta} E^{\alpha\beta} - \frac{1}{2} E^2 \right) - \frac{1}{2} a_3^\prime E^2.
\end{align}
Therefore, we can discard this term, if we are only interested in cases where $\varphi=t$, since, as for the operator $L_1^{(1,2)}$, it does not provide any new interesting information compared to the commonly discussed mimetic gravity scenarios. Note that this also applies to higher-order terms of the form $f(\Box\varphi) R_{\mu\nu} \varphi^\mu \varphi^\nu$.

\subsection{Quartic operators}
The relevant independent quartic operators which contribute to the EOM are 
\begin{align}
     L_1^{(2,2)} &= b_1 \varphi^{\mu\nu} \varphi_{\mu\nu} R, \quad L_2^{(2,2)}=b_2 \varphi^{\mu\nu} \varphi_{\mu\nu} \varphi^\alpha \varphi^\beta R_{\alpha\beta}, \quad L_3^{(2,2)}= b_3 (\Box\varphi)^2 R, \nonumber \\ L_4^{(2,2)} &=b_4 (\Box\varphi)^2 \varphi^\mu \varphi^\nu R_{\mu\nu}, \quad
    L_5^{(2,2)}=b_5 R_{\alpha\beta\mu\nu} \varphi^{\mu\alpha} \varphi^{\beta\nu}, \quad L_6^{(2,2)}=b_6 R_{\alpha\beta} \varphi^{\alpha\beta} \Box\varphi.
\end{align}
The form of $L_3^{(2,2)}$ and $L_4^{(2,2)}$ (accounting for the mimetic constraint) can be straightforwardly obtained from the previous results
\begin{align}
    L_3^{(2,2)} \vert_{N=1} =& b_3 E^2 \left(E_{ij} E^{ij} + \frac{1}{3} E^2 + \bar R \right) - \frac{2}{3} b_3^\prime E^3 , \\
    L_4^{(2,2)} \vert_{N=1}=& b_4 E^2 \left( \frac{1}{3} E^2 - E_{ij} E^{ij} \right)+ \frac{1}{3} b_4^\prime E^3 .
\end{align}
Further, by using the Gauss relation \eqref{eq:Gauss_equation} 
\begin{align}
    L_5^{(2,2)}\vert_{N=1}=& b_5 \left( \bar R_{\alpha\beta\mu\nu} + E_{\alpha\mu} E_{\beta\nu} - E_{\beta\mu} E_{\alpha\nu} \right)  E^{\alpha\mu} E^{\beta\nu} \nonumber \\
    =& b_5 \left[ 2 E \bar R_{\mu\rho} E^{\mu\rho} - 2 \bar R_{\nu\rho} E^\nu_\sigma E^{\sigma\rho} - \frac{1}{2} \bar R \left( E^2 - E_{\mu\nu} E^{\mu\nu} \right) + \left( E_{\alpha\mu} E_{\beta\nu} - E_{\beta\mu} E_{\alpha\nu} \right)  E^{\alpha\mu} E^{\beta\nu} \right],
\end{align}
where in the second step we have used that the 3-dimensional Riemann tensor can be decomposed in the Ricci scalar and Ricci tensor
\begin{align}
    \bar R_{\mu\nu\rho\sigma}= \bar R_{\mu\rho} h_{\nu\sigma} - \bar R_{\nu\rho} h_{\mu\sigma} - \bar R_{\mu\sigma} h_{\nu\rho} + \bar R_{\nu\sigma} h_{\mu\rho} - \frac{1}{2} \bar R \left( h_{\mu\rho} h_{\nu\sigma} - h_{\mu\sigma} h_{\nu\rho} \right).
\end{align}
Using the same argumentation as for the cubic operators, we obtain an effective field theory, with only three dof for $\varphi=t$. 

However, the situation changes for the remaining three higher-derivative operators
\begin{align}
    L_1^{(2,2)}\vert_{N=1} =& b_1 E_{\mu\nu} E^{\mu\nu} \left( \bar R + E_{\alpha\beta} E^{\alpha\beta} + E^2 \right) + 2 b_1 E_{\mu\nu} E^{\mu\nu}  \nabla_n E  \\
    L_2^{(2,2)}\vert_{N=1}=& - b_2 E_{\mu\nu} E^{\mu\nu}  E_{\alpha\beta} E^{\alpha\beta} - b_2 E_{\mu\nu} E^{\mu\nu} \nabla_n E \\
    L_6^{(2,2)}\vert_{N=1} =&  b_6 E (\bar R_{\alpha\beta} E^{\alpha\beta} + E_{\alpha\beta} E^{\alpha\beta} E + E^{\alpha\beta} \nabla_n E_{\alpha\beta}) \\
    =& b_6 \left(  E \bar R_{ij} E^{ij} + \frac{1}{2} E^2 E_{ij}E^{ij} - \frac{1}{2}  E^{ij} E_{ij}  \nabla_n E \right) - \frac{1}{2} b_6^\prime E E_{ij} E^{ij}
\end{align}
The time derivative of the extrinsic curvature cannot be rewritten anymore as a total derivative term. In the appendix \ref{sec:Appendix_Hamiltonian_analysis_Unitary_Gauge_L1}, we show explicitly that indeed $L_1^{(2,2)}$ leads to four dof even for fixed $\varphi=t$ from which one is an Ostrogradski ghost. So in order to obtain a theory with only three dof one has to introduce degeneracy conditions between the different operators in order to cancel the time derivatives of the extrinsic curvature. The degeneracy conditions for the quartic order are given by
\begin{align}
\label{eq:Degeneracy_condition_quartic}
  2 b_1 - b_2 - \frac{1}{2} b_6 = 0.
\end{align}
Finally, let us note that, in general, there may be more possible operators. However, choosing $\varphi=t$, the above operators form a complete set, from which we can construct all possible linear combinations of the 3-dim Ricci tensor and the extrinsic curvature
\begin{align}
\label{eq:ADM_quartic_order}
    \bar R E^2, \qquad \bar R E_{ij} E^{ij}, \qquad \bar R_{ij} E^{ij} E, \qquad \bar R_{i}^j E_j^k E_k^{i}.
\end{align}

\subsection{Quintic operators}
For the quintic operators linear in the curvature we have eleven possible operators
\begin{align}
   & L_1^{(3,2)}= c_1 (\Box\varphi)^3 R,\quad L_2^{(3,2)}= c_2 (\Box\varphi)^3 \varphi^\mu\varphi^\nu R_{\mu\nu}, \quad L_3^{(3,2)} = c_3 (\Box \varphi) \varphi_{\mu\nu} \varphi^{\mu\nu} R, \nonumber \\
    & L_4^{(3,2)}=c_4 (\Box\varphi) \varphi_{\mu\nu} \varphi^{\mu\nu} \varphi^\alpha \varphi^\beta R_{\alpha\beta}, \quad L_5^{(3,2)}= c_5 \varphi^{\mu\nu} \varphi^{\rho}_\mu \varphi_{\nu\rho} R, \quad L_6^{(3,2)}=c_6 \varphi^{\mu\nu} \varphi^{\rho}_\mu \varphi_{\nu\rho} \varphi^\alpha \varphi^\beta R_{\alpha\beta}, \nonumber \\
    & L_7^{(3,2)}= c_7 (\Box\varphi) \varphi^{\alpha\beta} \varphi^{\mu\nu} R_{\alpha\mu\beta\nu}, \quad  L_8^{(3,2)}= c_8 R_{\mu\nu} \varphi^{\mu\nu} \left( \Box\varphi \right)^2, \quad L_9^{(3,2)}=c_9 R_{\mu\nu} \varphi^{\mu\nu} \varphi_{\alpha\beta} \varphi^{\alpha\beta}, \nonumber \\ &  L_{10}^{(3,2)}=c_{10} R_{\mu\nu} \varphi^{\mu\alpha} \varphi_{\alpha\beta} \varphi^{\beta\nu}, \qquad L_{11}^{(3,2)} = c_{11} R_{\alpha\beta\mu\nu} \varphi^\beta \varphi^\nu \varphi^{\alpha\sigma} \varphi_{\sigma\gamma} \varphi^{\gamma\mu}.
    \label{eq:Quintic_linear_R}
\end{align}
Imposing the mimetic constraint we obtain
\begin{align}
   & L_1^{(3,2)}\vert_{N=1}= - c_1 E^3 \left( \bar R + E_{ij} E^{ij} + \frac{1}{2} E^2 \right) + \frac{1}{2} c_1^\prime E^4, \\
    & L_2^{(3,2)}\vert_{N=1}= -c_2 E^3 \left( \frac{1}{4} E^2 - E_{ij} E^{ij} \right) - \frac{1}{4} c_2^\prime E^4, \nonumber  \\
    & L_7^{(3,2)}\vert_{N=1}= - c_7 E \left( \bar R_{\alpha\beta\mu\nu} + E_{\alpha\mu} E_{\rho\nu} - E_{\beta\mu} E_{\alpha\nu} \right) E^{\alpha\mu} E^{\beta\nu}, \\
    & L_9^{(3,2)}\vert_{N=1} = - c_9 E_{ij} E^{ij} \left(  \bar R_{ij} E^{ij} + \frac{3}{4} E_{ij} E^{ij} E \right) + \frac{1}{4} c_9^\prime E_{ij} E^{ij} E_{kl} E^{kl}. 
\end{align}
Further, we get terms with time derivatives of the extrinsic curvature
\begin{align}
    L_3^{(3,2)}\vert_{N=1}=& - c_3 E E_{ij} E^{ij} ( \bar R + E_{kl} E^{kl} + E^2 ) - 2 c_3 E_{ij} E^{ij} E \nabla_n E, \\
    L_4^{(3,2)}\vert_{N=1}=& c_4 E ( E_{ij} E^{ij} )^2 + c_4 E E_{ij} E^{ij} \nabla_n E, \\
    L_5^{(3,2)}\vert_{N=1}=& - c_5 E_{ij} E^{ik} E_k^j ( \bar R + E_{lm} E^{lm} + E^2 ) - 2c_5 E_{ij} E^{ik} E_k^j \nabla_n E, \\
    L_6^{(3,2)}\vert_{N=1}=& c_6 E_{ij} E^{ik} E_k^j E_{lm} E^{lm} + c_6 E_{ij} E^{ik} E_k^j \nabla_n E, \\
     L_8^{(3,2)}\vert_{N=1} =& - c_8 E^2 \left( \bar R_{ij} E^{ij} + \frac{1}{2} E_{ij} E^{ij} E  \right) + c_8 E_{ij} E^{ij} E \nabla_n E + \frac{1}{2} c_8^\prime E^2 E_{ij} E^{ij}, \\
    L_{10}^{(3,2)}\vert_{N=1}=&\; - c_{10} \left( \bar R_{ij} + E_{ik} E^{k}_j\right) E^{il} E_{lm} E^{mj} - c_{10} E^{ik} E_{kl} E^{lj} \nabla_n E_{ij}, \\
    L_{11}^{(3,2)}\vert_{N=1}=& c_{11} E_{im} E_j^m  E^{ik} E_{kl} E^{lj} + c_{11}  E^{ik} E_{kl} E^{lj} \nabla_n E_{ij}. 
\end{align}
To remove the time derivatives we need three degeneracy conditions
\begin{align}
     2 c_3 - c_4 -c_8 =0, \quad 2 c_5 - c_6=0, \quad c_{11} - c_{10} =0.
\end{align}
Note, that in comparison to \cite{Gorji:2017cai} we have added four operators $L_8^{(3,2)}$ - $L_{11}^{(3,2)}$, which were missing there. On the other hand, several operators from the aforementioned reference are obsolete for mimetic gravity theories. Moreover notice that one may construct more possible operators. 
But, restricting to $\varphi=t$, the operators $L_1^{(3,2)}$ - $L_{11}^{(3,2)}$ already form a complete set of all 3-dim operators which are linear in the curvature
\begin{align}
     & \bar R E^3, \qquad \bar R E E_{ij} E^{ij}, \qquad \bar R E_{ij} E^{jk} E_k^i,\qquad \bar R_{ij} E^{ik} E_k^j E, \nonumber \\ &
    \bar R_{ij} E^{ij} E^2, \qquad \bar R_{ij} E^{ij} E_{mn} E^{mn}, 
     \qquad \bar R_{ij} E^i_k E^k_l E^{lj}.
\end{align}

\subsubsection*{Quadratic in the curvature}

As in \cite{Gorji:2017cai} one could also discuss coupling terms between the scalar field and the curvature which are quadratic in the curvature. Indeed it is straightforward to see that the operators
\begin{align}
\label{eq:Quadratic_operator_quintic}
   \Box\varphi \left( \frac{1}{2} R + R_{\mu\nu} \varphi^\mu \varphi^\nu \right)^2 \Big\vert_{N=1} &= - \frac{1}{4} E \left( \bar R - E_{ij} E^{ij} + E^2 \right)^2, \\
   \varphi^\alpha_\sigma R_{\alpha\epsilon} R_\lambda^\sigma \varphi^\epsilon \varphi^\lambda  \Big\vert_{N=1} &= - E^{\alpha}_{\sigma} \left( D_\nu E^{\nu\sigma} - D^\sigma E \right) \left( D_\mu E^\mu_\alpha - D_\alpha E \right).
\end{align}
lead to three dof in the homogeneous limit $\varphi=t$. We can see that in addition to the combination of the 3-dim curvature $\bar R_{ij}$ and the extrinsic curvature $E_{kl}$ we also obtain spatial covariant derivatives of the extrinsic curvature $E_{ij}$. However, we restrict ourselves to the construction of all possible independent combinations of $R_{ij}$ and $E_{kl}$ without spatial covariant derivatives of the extrinsic curvature which are commonly used to construct effective field theories 
\begin{align}
\label{eq:R_quadratic}
    \bar R^2 E, \qquad \bar R_{ij} E^{ij} \bar R, \qquad \bar R_{ij} \bar R^{ij} E, \qquad \bar R_{ij} \bar R^{jk} E_k^i.
\end{align}
In contrast to the previous parts we do not construct all the covariant operators and then derive the degeneracy condition but instead construct directly the four spatial operators \eqref{eq:R_quadratic} from the covariant operators for simplicity. 

From the subsection \ref{subsec:Relation_curvature} it is straightforward to see that we can construct the 3-dim curvature out of
\begin{align}
    \bar R =& R + K^2 - K_{\mu\nu} K^{\mu\nu} - 2 R_{\alpha\beta} n^\alpha n^\beta, \\
    \bar R_{\mu\nu}=& R_{\mu\alpha\nu\beta} n^\alpha n^\beta - K K_{\mu\nu} + K_{\mu\sigma} K^\sigma_\nu + h^\alpha_\mu h^\beta_\nu R_{\alpha\beta}.
\end{align}
Using these relations we obtain
\begin{align}
    \bar R^2 E =& -\Box\varphi \left( R + (\Box\varphi)^2 - \varphi_{\mu\nu} \varphi^{\mu\nu} - 2 R_{\alpha\beta} \varphi^\alpha \varphi^\beta \right)^2 \Big\vert_{N=1}, \\
    \bar R_{ij} E^{ij} \bar R =& \left( - R_{\mu\mu} \varphi^{\mu\nu} - R_{\mu\alpha\nu\beta} \varphi^{\mu\nu} \varphi^\alpha\varphi^\beta + \Box\varphi \varphi_{\mu\nu} \varphi^{\mu\nu} - \varphi_{\mu\sigma} \varphi^\sigma_\nu \varphi^{\mu\nu} \right) \nonumber \\ &\times \left(  R + (\Box\varphi)^2 - \varphi_{\mu\nu} \varphi^{\mu\nu} - 2 R_{\alpha\beta} \varphi^\alpha \varphi^\beta \right) \Big\vert_{N=1}, \\
    \bar R_{ij} \bar R^{ij} E =& -\Box \varphi \Big[ R_{\mu\nu} R^{\mu\nu} + 2 R_{\alpha\nu} \varphi^\alpha R^{\beta\nu} \varphi_\beta + ( R_{\alpha\beta} \varphi^\alpha \varphi^\beta)^2 + 2 R^{\mu\nu} \Big(  R_{\mu\alpha\nu\beta} \varphi^\alpha\varphi^\beta - \Box\varphi \varphi_{\mu\nu} + \varphi_{\mu\sigma} \varphi^\sigma_\nu \Big) \nonumber \\
    & + \left( R_{\mu\alpha\nu\beta} \varphi^\alpha\varphi^\beta - \Box\varphi \varphi_{\mu\nu} + \varphi_{\mu\sigma} \varphi^\sigma_\nu \right)^2\Big] \Big\vert_{N=1}, \\
    \bar R_{i}^j \bar R^{k}_j E_k^i =& - R_{\alpha\beta} R^{\beta\sigma} \varphi^\alpha_\sigma  - R_{\alpha\delta} R^\sigma_\epsilon \varphi^\alpha_\sigma \varphi^\delta \varphi^\epsilon - 2 \varphi^\alpha_\sigma  R_{\alpha\beta} \Big( R^{\beta\delta\sigma\epsilon} \varphi_\delta \varphi_\epsilon - \Box \varphi \varphi^{\beta\sigma} + \varphi^\beta_\epsilon \varphi^{\epsilon\sigma} \Big) \nonumber \\
    & - \varphi^\alpha_\sigma \left( R_{\alpha\beta} + R_{\alpha\rho \beta\delta} \varphi^\rho \varphi^\delta - \Box\varphi \varphi_{\alpha\beta} + \varphi_{\alpha\delta} \varphi^{\delta}_\beta \right) \left( R^{\beta\delta\sigma\epsilon} \varphi_\delta \varphi_\epsilon - \Box \varphi \varphi^{\beta\sigma} + \varphi^\beta_\epsilon \varphi^{\epsilon\sigma}  \right) \Big \vert_{N=1}.
\end{align}

\section{Hamiltonian analysis for general field configurations}
\label{sec:Hamiltonian_analysis}
In the previous section we have analyzed the higher-derivative operators in which the time flow is identified with the mimetic scalar field, $\varphi=t$. As discussed in \cite{Blas:2009yd,Deffayet:2015qwa,Langlois:2015skt} the choice of a homogeneous scalar field can alter the number of dof. Therefore, we perform the Hamiltonian analysis for the previously discussed operators for a generic scalar field $\partial_i\varphi \neq 0$ to analyze possible effects on the number of the dof.

\subsection{Notation and auxiliary variables}

To perform the Hamiltonian analysis and simplify the notation we introduce the following auxiliary variables
\begin{align}
    A_\star =& n^\mu \nabla_\mu \varphi = \frac{1}{N} \left( \dot\varphi - N^i D_i \varphi \right), \\
    V_\star=& n^\mu n^\nu \nabla_\mu \nabla_\nu \varphi= \frac{1}{N} \left( \dot A_\star - N^i D_i A_\star - N a_i D^i \varphi \right), \\
    C_{ij}=& D_i D_j \varphi - A_\star K_{ij}, \\
    C_i =& n^\nu h^\mu_i \nabla_\mu \nabla_\nu \varphi = D_i A_\star - K_i^j D_j \varphi.
\end{align}
Some further useful relations are
\begin{align}
    \Box \varphi  =& C_{i}^{i} - V_\star = D^i D_i \varphi- A_\star K - V_\star, \\
    \varphi_{\mu\nu} \varphi^{\mu\nu}=& C_{ij} C^{ij} - 2 C_i C^i + V_\star^2 \nonumber \\ =& \left( D_i D_j \varphi - A_\star K_{ij} \right) \left( D^i D^j\varphi - A_\star K^{ij} \right) - 2 \left( D_i A_\star - K_i^j D_j\varphi \right) \left( D^i A_\star - K^{ik} D_k \varphi \right) + V_\star^2.
\end{align}
For the cubic operators we obtain
\begin{align}
\label{eq:2_1_2}
    L_2^{(1,2)} =& a_2 \left(C_i^i - V_\star \right) \left[ \bar R + K_{ij} K^{ij} +K^2 - 2 D_i a^i - 2 a_i a^i + 2 \nabla_n K \right], \\
    L_3^{(1,2)} =& a_3 \left( C_i^i - V_\star \right) \left[ _\bot R_{ij} D^i \varphi D^j \varphi - 2 _\bot R_{in} A_\star D^i\varphi + R_{nn} A_\star^2 \right] \nonumber \\
     =& a_3 \left( C_i^i - V_\star \right)  \Big[ \bar R_{ij} D^i\varphi D^j\varphi + K K_{ij} D^i\varphi D^j\varphi + \nabla_n K_{ij} D^i\varphi D^j\varphi - D_i a_j D^i\varphi D^j\varphi  \nonumber \\ & - a_i a_j D^i\varphi D^j\varphi - 2 A_\star D^i\varphi \left( D_j K^j_i - D_i K\right) - A_\star^2 K_{ij} K^{ij} + A_\star^2 D_i a^i  + A_\star^2 a_i a^i - A_\star^2 \nabla_n K \Big].
     \label{eq:3_1_2}
\end{align}
As discussed before, we do not consider $L_1^{(1,2)}$ since it does not provide any new information (see subsection \ref{subsec:Cubic_operators} ). 
For later purposes, note:
\begin{align}
    D_i a_j + a_i a_j = \frac{1}{N} D_i D_j N.
\end{align}
Similar to \cite{BenAchour:2016fzp} it is convenient to split $K_{ij}$ into different components by projecting orthogonal to $D_i \varphi$, namely
\begin{align}
    K_{ij} =&  P^k_i P^l_j K_{kl} + 2 \frac{D^k \varphi  P^l_{(j} D_{i)} \varphi K_{lk}}{D_m \varphi D^m\varphi}  + E \frac{D_i\varphi D_j\varphi }{D_k \varphi D^k\varphi} \nonumber \\
    \equiv &  _\bot \hat  K_{ij} + \hat K_{ij} + E \frac{D_i\varphi D_j\varphi}{D_k\varphi D^k\varphi}
\end{align}
where we have introduced the projection operator
\begin{align}
    P^k_i = h^k_i - \frac{D^k\varphi D_i \varphi}{D_l \varphi D^l \varphi}.
\end{align}
Further, the matrices have the properties
\begin{align}
    _\bot \hat {K}_{ij} h^{ij} =&  P^k_i P^l_j K_{kl} h^{ij}= F, \qquad _\bot \hat  K_{ij} D^i \varphi = 0, \nonumber \\
    \hat K_{ij} h^{ij} =& 2  \frac{D^k \varphi  P^l_{(j} D_{i)} \varphi K_{lk}}{D_m \varphi D^m\varphi} h^{ij}=0, \qquad \hat K_{ij} D^i\varphi = K_{lk} P^l_j D^k\varphi.
\end{align}
From \eqref{eq:2_1_2} and \eqref{eq:3_1_2} we can see that in the case of the two cubic operators two components of the extrinsic curvature acquire a time derivative
\begin{align}
    E = K_{ij} \frac{D^i \varphi D^j \varphi}{D_k \varphi D^k\varphi} , \qquad V_E =\nabla_n E, \\
    F= \left( h^{ij} - \frac{D^i \varphi D^j\varphi}{D_k \varphi D^k\varphi} \right) K_{ij}, \qquad V_F=\nabla_n F.
\end{align}
For later purposes, we will also need the relation
\begin{align}
    D^i \varphi D^j\varphi \nabla_n K_{ij} =& D_k \varphi D^k\varphi V_E - 2 D_j \varphi K^{ij} \left( D_i A_\star - K_i^k  D_k \varphi +  A_\star a_i \right) \nonumber \\ & + \frac{2 K_{ij} D^i\varphi D^j\varphi}{D_k \varphi D^k\varphi} \left( D^l\varphi D_l A_\star - K_{kl} D^k\varphi D^l\varphi + a_l D^l\varphi A_\star \right) \nonumber \\
    =& D_k \varphi D^k\varphi V_E - 2 D_j \varphi \hat K^{ij} \left( D_i A_\star - \hat K^k_i D_k \varphi  \right) - 2 D_j \varphi \hat K^{ij} a_i A_\star.
\end{align}

\subsection{Hamiltonian analysis: cubic order}
\label{subsec:Hamiltonian_analysis_cubic_order}
Using the splitting procedure from above and some partial integration (up to total derivatives) we can rewrite the two cubic operators, $L_3^{(1,2)}$ and $L_2^{(1,2)}$, as 
\begin{align}
    L^{(1,2)} =& a_3 \left(C_k^k - V_\star \right) \Big[ \bar R_{ij} D^i\varphi D^j\varphi + ( E + F) D_i \varphi D^i \varphi E  + 2 A_\star D_i D_j \varphi _\bot \hat K^{ij} \nonumber \\ & 
    - 2 A_\star  E D^i \varphi D_j \left( \frac{D_i \varphi D^j \varphi}{D_m \varphi D^m\varphi} \right)  + 2 A_\star D_i F D^i \varphi  - A_\star^2 \Big( {_\bot \hat K_{ij} } {_\bot \hat K^{ij} }  + \hat K_{ij} \hat K^{ij} + E^2\Big) \nonumber \\ & - A_\star^2 V_F +V_E \left( D_k \varphi D^k\varphi - A_\star^2 \right) - 2 D_j \varphi \hat K^{ij} \left( D_i A_\star - \hat K^k_i D_k\varphi \right) \Big]
    \nonumber \\
    & +  a_2 \left(C_k^k - V_\star \right) \Big[ \bar R + _\bot{\hat K^{ij}} _\bot{ \hat K_{ij}} +\hat K_{ij} \hat K^{ij} + E^2 + \left( E + F \right)^2 + 2 V_E + 2 V_F  \Big]
    \nonumber \\ & + 2 \hat K^{ij} D_i \Big[ a_3 \left(C_k^k - V_\star \right) D_j \varphi  A_\star \Big] - 2 D_k D^k \left[ a_2 \left( C_i^i - V_\star \right) \right] \nonumber \\ & + D_i D_j \Big[ a_3 \left(C_k^k - V_\star \right) \left( D^i \varphi D^j \varphi - A_\star^2 h^{ij}  \right) \Big] 
\end{align}
Introducing new auxiliary variables we can rewrite the action as
\begin{align}
\label{eq:action_cubic_operators_Hamiltonian}
    S = \int \md^3x\,\md t\; &   \sqrt{h} \Big[ N L^{(1,2)} +  N \lambda \left( A_\star^2 - D_i D^i\varphi - 1\right)+ \mu  \left( N A_\star - \dot \varphi + N^i D_i \varphi \right) - N D_i\left( \zeta D^i\varphi \right)  \nonumber \\ &   + \zeta \left( N V_\star - \dot A_\star + N^i D_i A_\star \right) + N _\bot \hat{ \tilde \Lambda}^{ij}  \left( _\bot \hat{\tilde B}_{ij} - _\bot \hat{ \tilde K}_{ij} \right)  + N \hat{\Lambda}^{ij} \left(\hat B_{ij} - \hat K_{ij} \right) \nonumber \\ & 
    + N \Sigma_E \left( E - \frac{K_{ij} D^i\varphi D^j\varphi}{D_k \varphi D^k\varphi} \right) + \frac{1}{3} N \Sigma_F  \left( F - \left( h^{ij} - \frac{D^i\varphi D^j\varphi}{D_k \varphi D^k\varphi} \right) K_{ij} \right)  \Big],
\end{align}
where we have split, further, $_\bot \hat K_{ij}$ into its trace $F$ and traceless part $_\bot \hat{\tilde K}_{ij}$. The first term in the action comes from the two cubic operators and the second one from the mimetic constraint. The other terms in the action are Lagrange multiplier terms to fix the introduced auxiliary variables. For simplicity, we will assume in the following that $a_2$ and $a_3$ are constant.

\subsubsection{Analysis of the constraints}
The canonical conjugate momenta obtained from \eqref{eq:action_cubic_operators_Hamiltonian} are all primary constraints. The non-trivial are given by
\begin{align}
    \bar P_E =&  \sqrt{h} \left( D_k D^k \varphi - A_\star ( F+E) - V_\star \right) \left[  a_3 \left( D_k \varphi D^k \varphi - A_\star^2\right) + 2 a_2 \right] - P_E \approx 0 \\
    \bar P_F =&  \sqrt{h} \left( D_k D^k \varphi - A_\star ( F+E) - V_\star \right) \left[ - a_3 A_\star^2 + 2 a_2 \right] - P_F \approx 0 \\
    \bar p_\varphi =& \sqrt{h} \mu + p_\varphi \approx 0, \label{eq:Primary_constraint_p_f} \\
    \bar p_\star =& \sqrt{h} \zeta + p_\star \approx 0, \label{eq:Primary_constraint_p_star} \\
    _\bot \hat{ \bar \pi}_{ij} = & \frac{1}{2} \sqrt{h} _\bot \hat{\tilde \Lambda}_{ij} + _\bot \hat{\tilde \pi}_{ij} \approx 0, \qquad \hat {\bar \pi}_{ij}= \frac{1}{2} \sqrt{h} \hat \Lambda_{ij} + \hat{\pi}_{ij} \approx0, \nonumber \\
    \bar \pi_E =& \frac{1}{2} \sqrt{h} \Sigma_E + \pi_E\approx 0, \qquad  
    \bar \pi_F = \frac{1}{6} \sqrt{h} \Sigma_F + \pi_F \approx 0, \label{eq:Primary_constraint_Pi}  
\end{align}
where "$\approx$" denotes weak equalities, which are only valid on the constraint surface, while "$=$" denotes strong equalities which are valid in the whole phase-space \cite{Dirac:1958sq}.
To clarify the notation we have 28 canonical conjugate pairs
\setcounter{MaxMatrixCols}{20}
\begin{align}
\nonumber
    \begin{pmatrix}
    N & N^i & h_{ij} & \varphi & \lambda & A_\star & \mu & \zeta & V_\star & _\bot \hat{\tilde \Lambda}_{ij} & \hat \Lambda_{ij} & _\bot \hat{\tilde B}_{ij} & \hat B_{ij} & \Sigma_E & E & \Sigma_F & F \\
    \pi_N & \pi^i & \pi_{ij} & p_\varphi & p_\lambda & p_\star & p_\mu & p_\zeta & P_{V_\star} & (\hat{\tilde P}_\Lambda )_{ij} & (\hat P_\Lambda)_{ij} & (\hat{\tilde P}_B)_{ij} & (\hat P_B)_{ij} & P_{\Sigma_E} & P_{E} & P_{\Sigma_F} & P_F
    \end{pmatrix}
\end{align} 
The extended Hamiltonian can be written as
\begin{align}
    H_T = \int \md^3x \left[ N \mathcal{H} + N^i \mathcal{H}_i  + \sum_A u_A P_A \right],
\end{align}
where the sum runs over all momenta (primary constraints). Further, the Hamiltonian and momentum constraint are equal to 
\begin{align}
  \mathcal{H} = &\frac{a_3 P_F}{a_3 A_\star^2 - 2 a_2}  \Big[ \bar R_{ij} D^i\varphi D^j\varphi + E ( E + F) D_i \varphi D^i \varphi   + 2 A_\star {_\bot}\hat{\tilde B}^{ij}  D_i D_j \varphi + \frac{2}{3} A_\star F D_i D^i \varphi  - 2 A_\star   E D^i \varphi \nonumber \\ & \times  D_j \left( \frac{D_i \varphi D^j \varphi}{D_m \varphi D^m\varphi} \right)  + 2 A_\star  D^i \varphi D_i F  - A_\star^2 \Big( {_\bot \hat{ \tilde B}_{ij} } {_\bot \hat{ \tilde B}^{ij} } + \frac{1}{3} F^2  + \hat B_{ij} \hat B^{ij} + E^2\Big)  - 2 D_j \varphi \hat B^{ij}  \nonumber \\ & \times \left( D_i A_\star - \hat B^k_i D_k\varphi \right) \Big] + \frac{a_2 P_F}{a_3 A_\star^2 - 2 a_2} \Big[ \bar R + \left( E + F \right)^2 + _\bot{\hat{\tilde B}_{ij} } {_\bot \hat{\tilde B}^{ij} } + \hat B_{ij} \hat B^{ij} + \frac{1}{3} F^2 + E^2 \Big] \nonumber \\
  & + 2 \sqrt{h}  \hat B^{ij}  D_i \Big[ \frac{a_3}{a_3 A_\star^2 - 2 a_2} \frac{P_F}{\sqrt{h}} A_\star D_j \varphi  \Big] + \sqrt{h} D_i D_j \Big[  \frac{1}{a_3 A_\star^2 - 2 a_2} \frac{P_F}{\sqrt{h}} \big[ a_3 \left( D^i \varphi D^j \varphi - A_\star^2 h^{ij}  \right)  \nonumber \\
  & - 2 a_2 h^{ij} \big] \Big] + p_\varphi A_\star + 2 _\bot\hat{\tilde \pi}^{ij} {_\bot} \hat{\tilde B}_{ij} +  2 \hat \pi^{ij} \hat B_{ij} + 2 \pi_E E + 2 \pi_F F + p_\star V_\star - D_i \left( p_\star D^i \varphi \right) \nonumber \\ &
  + \lambda \sqrt{h} \left( - A_\star^2 + D_i \varphi D^i\varphi + 1 \right), \\
  \mathcal{H}_i =& - 2 D^j \pi_{ij} + p_\varphi D_i \varphi + p_\star D_i A_\star + P_{E} D_i E + P_F D_i F,
  \label{eq:Diffeomorphism_constraint}
\end{align}
Note, that the other primary momenta are all weakly zero. Therefore, we could add them to the momentum constraint. It is then straightforward to show that $\mathcal{H}_i$ is weakly equal to first-class constraints corresponding to the time-independent spatial transformations.

As a next step, we have to check the time evolution of all the primary constraints. The conservation of $\bar p_\varphi$, $\bar p_\star$, $_\bot \hat{\bar \pi}_{ij}$, $\hat{\bar \pi}_{ij}$, $\pi_E$ and $\bar \pi_F$ fixes the Lagrange parameters $u_\mu$, $u_\zeta$, $(\hat{\tilde u}_\Lambda)_{ij}$, $(\hat u_\Lambda)_{ij}$, $u_{\Sigma_E}$ and $u_{\Sigma_F}$ and vice versa. 

The conservation of $\pi_N$ and $\pi^i$ yields the usual Hamiltonian and momentum constraint
\begin{align}
    - \dot \pi_N = \mathcal{H}\approx 0, \qquad - \dot \pi^i = \mathcal{H}^i \approx 0.
\end{align}
As discussed before, $\mathcal{H}_i$ is weakly equal to first-class constraints. On the other hand, it is very involved to show that $\mathcal{H}$ together with $\pi_N$ are weakly equal to first-class constraints responsible for the time transformations. However, it is natural to expect it due to the diffeomorphism invariance of the starting theory and we will assume it in the following (see \cite{Crisostomi:2017ugk, Takahashi:2017pje,Langlois:2015skt} for related discussions).
The mimetic constraint is given by the conservation of $p_\lambda$, namely
\begin{align}
    - \frac{1}{ N } \dot p_\lambda= \sqrt{h} \left( - A_\star^2 + D_i \varphi D^i\varphi +1 \right)  \equiv C_\lambda \approx 0.
    \label{eq:Mimetic_constraint}
\end{align}
The conservation of the other primary constraint leads to 
\begin{align}
     \dot P_{V_\star} \approx & \sqrt{h} \left( - a_3 A_\star^2 + 2 a_2 \right) u_F + \sqrt{h} \left( - a_3 + 2 a_2 \right) u_E + \{P_{V_\star}, \int \md^3x\, N \mathcal{H} \} \approx 0, \\
     \dot{\bar P}_F \approx & \sqrt{h} a_3 \left(A_\star - A_\star^3 \right) u_E + \sqrt{h} \left(a_3 A_\star^2 - 2 a_2 \right) u_{V_\star} + \{\bar P_F,\int \md^3x N \mathcal{H} \} \approx 0, \\
    \dot{\bar P}_E \approx & \sqrt{h} a_3 \left(  A_\star^3 - A_\star \right)  u_F + \sqrt{h} \left( a_3 - 2 a_2 \right) u_{V_\star} + \{ \bar P_E, \int \md^3x N \mathcal{H} \}  \approx 0,
\end{align}
where $\{.\,,.\}$ denotes the Poisson bracket. The equations above are not all independent and two of them can be used to fix $u_F$ and $u_E$ in terms of $u_{V_\star}$
\begin{align}
\label{eq:Lagrange_u_F}
    u_F \approx & \frac{1}{\sqrt{h} a_3} \frac{1}{A_\star (A_\star^2 - 1)} \Big[ \sqrt{h} \left( 2a_2 - a_3 \right) u_{V_\star} - \{ \bar P_E, \int \md^3x N \mathcal{H} \} \Big],  \\
    u_E \approx &  \frac{1}{\sqrt{h} a_3} \frac{1}{A_\star( A_\star^2 - 1)} \Big[ \sqrt{h} \left( a_3 A_\star^2 - 2 a_2 \right) u_{V_\star} + \{ \bar P_F, \int \md^3x N \mathcal{H} \} \Big].
    \label{eq:Lagrange_u_E}
\end{align}
While $A_\star \approx 0$ is inconsistent with the mimetic constraint \eqref{eq:Mimetic_constraint}, we have to note that these relations only hold for $A_\star^2 \not \approx 1$, which implies $\varphi \not \approx t$.
Additionally, we obtain one secondary constraint, which is given by
{\allowdisplaybreaks
\begin{align}
   N C_E \equiv & ( 2 a_2 - a_3 ) \{ \bar P_F, \int \md^3x N \mathcal{H} \} - \left( 2 a_2 - a_3 A_\star^2 \right) \{ \bar P_E, \int \md^3x N \mathcal{H} \} \nonumber \\ 
    & + a_3 ( A_\star^3 - A_\star ) \{ P_{V_\star}, \int \md^3x N \mathcal{H} \} \nonumber \\
    \approx & N \frac{P_F}{a_3 A_\star^2 - 2 a_2} \Big\{ \left( 2 a_2 - a_3 \right) a_3 \Big[ 2 A_\star V_\star + E D_i \varphi D^i\varphi + \frac{2}{3} A_\star D_k D^k \varphi - \frac{2}{3} F A_\star^2\Big] \nonumber \\
    & + ( 2 a_2 - a_3) a_2 \left( 2 E + \frac{8}{3} F \right) + ( 2 a_2 - a_3 A_\star^2) a_3 \Big[ - ( 2 E + F) D_i \varphi D^i \varphi + 2 A_\star D^i \varphi  \nonumber \\ & \times D_j \left( \frac{D_i \varphi D^j\varphi}{D_k \varphi D^k\varphi} \right) + 2 A_\star^2 E \Big] - ( 2 a_2 - a_3 A_\star^2 ) a_2 ( 4 E + 2F )   \Big\} - N ( 2 a_2 - a_3 ) a_3 \nonumber \\
    & \times D_i \left( \frac{P_F}{a_3 A_\star^2 - 2a_2} D^i \varphi A_\star \right) - N a_3 A_\star D^i\varphi D_i \varphi p_\star +2 N  ( 2a_2 -a_3) \pi_F \nonumber \\
    & - 2 N ( 2 a_2 -a_3 A_\star^2 ) \pi_E.
\end{align}}
The conservation of $(\hat{\tilde P}_B)_{ij}$, $(\hat P_B)_{ij}$ yields 
\begin{align}
     \frac{1}{N} \frac{\md }{\md t}({\hat{\tilde P}}_B)_{ij} \approx &    \frac{P_F}{a_3 A_\star^2 - 2 a_2} \Big[ 2 \left( a_3 A_\star^2 - a_2 \right){ _\bot \hat{\tilde B}_{ij}} - 2 a_3 A_\star \left( P^k_i P^l_j - \frac{1}{3} h_{ij} P^{kl} \right) D_k D_l \varphi  \Big] - 2  _\bot\hat{\tilde \pi}_{ij} \nonumber \\
      \equiv & (_\bot \hat{\tilde C}_B)_{ij} \approx 0, \\
     \frac{1}{N} \frac{\md}{\md t}( { \hat P}_B)_{ij} \approx &  \frac{P_F}{a_3 A_\star^2 - 2 a_2} \Big[ 2 ( a_3 A_\star^2 - a_2 ) \hat B_{ij} +  a_3  P^k_{(j}  D_{i)} \varphi  D_{k} A_\star  - 2 a_3 D_{(i} \varphi \hat B^k_{j)} D_k \varphi \Big] \nonumber \\ 
     & - 2 a_3  \sqrt{h}   \frac{ P^k_{(j} D_{i)} \varphi D^l \varphi}{D^m \varphi D_m \varphi } D_{(k} \left( \frac{1}{\sqrt{h}} \frac{P_F}{a_3 A_\star^2- 2 a_2} D_{l)} \varphi \right)- 2 \hat \pi_{ij} \equiv (\hat C_B)_{ij} \approx 0
\end{align}
As a next step, we have to check the conservation of the secondary constraints. While $(_\bot \hat{\tilde C}_B)_{ij}$ and $(\hat C_B)_{ij}$ fixes the Lagrange parameter $(_\bot \hat{\tilde u}_B)_{ij}$ and $(\hat u_B)_{ij}$, the conservation of $C_\lambda$ yields a tertiary constraint
\begin{align}
    \frac{1}{N}\dot C_\lambda = - 2 \sqrt{h} A_\star V_\star  +  2 \sqrt{h} D^i \varphi D_i A_\star  - 2 \sqrt{h} D_i \varphi D^i \varphi E \equiv C_\lambda^{(2)} \approx 0. \label{eq:C_lambda_2_cubic}
\end{align}
Using \eqref{eq:Lagrange_u_F} and \eqref{eq:Lagrange_u_E} the conservation of $C_E$ yields
\begin{align}
    \dot C_E(y) \approx & - 4 \sqrt{h} N a_2 \lambda D_i \varphi D^i \varphi + N J( h_{ij}, \pi_{ij}, ... ) + \int \md^3x\, u_{V_\star} \Big[\frac{1}{a_3} \frac{1}{A_\star^3-A_\star} \Big( (2 a_2 -a_3) \nonumber \\
    & \times \{ C_E(y) , \bar P_F \}  + ( a_3 A_\star^2 - 2 a_2 ) \{ C_E(y), \bar P_E \} \Big) + \{ C_E(y), P_{V_\star} \} \Big].
    \label{eq:Conservation_C_E}
\end{align}
It can be used to fix $u_{V_\star}$, which will explicitly depend on $\lambda$.
The conservation of $C_\lambda^{(2)}$ leads to a new constraint
\begin{align}
    \dot C_\lambda^{(2)} \approx & \left( - 2 \sqrt{h} A_\star + 2 \sqrt{h} \frac{a_3 A_\star^2 - 2 a_2}{a_3 A_\star}\right) u_{V_\star} + 2 \frac{D_k \varphi D^k\varphi}{a_3 ( A_\star^3 - A_\star )} \{ \bar P_F , \int \md^3x N \mathcal{H} \} \nonumber \\ 
    & + \{ C_\lambda^{(2)}, \int d^3x\, N \mathcal{H} \} \equiv N C_\lambda^{(3)} \approx 0.
\end{align}
Since it depends on $u_{V_\star}$ the new constraint depends explicitly on $\lambda$. Therefore, the conservation of the new constraint $C_\lambda^{(3)}$ will fix $u_\lambda$ and the chain of constraints stops here.

At the end, we have the standard eight first-class constraints and 32 
second-class constraints. Since we have 56 canonical variables we end up with four degrees of freedom. However, one can show that
\begin{align}
    \{ p_\lambda, C_\lambda^{(3)} \} \propto D_i \varphi D^i \varphi,
\end{align}
while $p_\lambda$ weakly commutes with all the other constraints. Consequently, we can observe that similarly to \cite{Zheng:2018cuc} the Dirac matrix is only invertible if and only if (iff) $\partial_i \varphi \not\approx 0$. 
Equivalently, we can see that for this case the constraint $C_\lambda^{(3)}$ is not conserved anymore. Therefore, for homogeneous scalar fields $\partial_i \varphi \approx 0$ the analysis has to be redone, starting from checking the time conservation of the primary constraints. Eventually, we obtain that we have to impose further constraints. 
In general, we have to distinguish the two different possible branches of solutions with $\partial_i \varphi \approx 0$ and $\partial_i \varphi  \not\approx 0$ (see \cite{Gomes:2017tzd} for a similar discussion in the context of cuscuton models).

\subsubsection{Homogeneous field configuration}
Let us now analyze the case of a homogeneous field $\partial_i \varphi \approx 0$ more explicitly. We will only outline the most important results and refer to \cite{Zheng:2018cuc}, and to the appendix \ref{sec:Degeneracy_1_3_model}, where we discuss the same feature in detail for a simpler mimetic gravity model. 

For the case $\partial_i \varphi\approx 0$ we obtain $\varphi\approx f(t)$. Combining with the mimetic constraint we introduce the constraint
\begin{align}
    C_t = \varphi-t \approx 0.
\end{align}
The propagation of the constraint yields the secondary constraint
\begin{align}
    \dot C_t = -1 + N A_\star \equiv C_t^{(2)} \approx 0.
\end{align}
These two constraints are second class and reduce the two first class constraints $\pi_N$ and $\mathcal{H}$ to second class due to 
\begin{align}
    \{\pi_N , C_t^{(2)}  \} = - A_\star, \qquad \{C_t, \mathcal{H} \} \approx A_\star.
\end{align}
In this sense we have just the usual gauge fixing conditions for the unitary gauge in mimetic gravity theories. In fact, for "standard" mimetic gravity model, as the original mimetic matter theory \cite{Chamseddine:2013kea}, these two constraints are just unitary gauge fixing conditions. However, as discussed, we can see that imposing only the two second-class constraints $C_t$ and $C_t^{(2)}$ as gauge-fixing condition leads to a singular Dirac matrix, since $p_\lambda$ now commutes weakly with all the constraints. 
On the other hand, for $\varphi\approx t$  $C_\lambda^{(3)}$ does not depend anymore on $\lambda$ and, therefore, its time conservation does not fix $u_\lambda$. Instead, we have to impose further constraints, since otherwise the mimetic constraint would not be conserved in time.
Indeed it is straightforward to see that the time conservation of $C_\lambda^{(3)}$ leads to two additional second-class constraints $C_\lambda^{(4)}$ and $C_\lambda^{(5)}$, whereby the latter depends explicitly on $\lambda$ and the chain of constraint stops there. This mechanism is the same as discussed in \cite{Zheng:2018cuc} and for a particular model in the appendix \ref{sec:Degeneracy_1_3_model}. 

Due to the two additional second-class constraints the number of degrees of freedom is reduced by one. Imposing the conditions $C_t$ and $C_t^{(2)}$ cannot be considered just as gauge-fixing conditions, since they introduce further constraints. Instead, as outlined before, we can note that there are two different branches of models with $\partial_i \varphi \approx 0$ and $\partial_i \varphi \not \approx 0$, which are not equivalent. Instead, the transition from one to the other branch is singular and not well defined. By imposing the homogeneous profile for the scalar field we pick up explicitly one of the two possible branches of solutions.

Note, further that the first branch with $\partial_i \varphi \not \approx 0$ is only well defined if this condition holds at all times, since otherwise the EOM breaks down in the singular point where $\partial_i \varphi(x) \approx 0$ \cite{Gomes:2017tzd}. It is, however, beyond the scope of this paper to check it explicitly.

On the other hand, the branch of solutions with $\varphi \approx t$ obtained by imposing the constraints $C_t$ and $C_t^{(2)}$ is well defined. Note that this is equivalent to directly imposing the homogeneous scalar field in the action, as done in section \ref{sec:Higher_Derivatives}. It is an explicitly Lorentz breaking theory with three degrees of freedom (no Ostrogradski ghost). Alternatively, it might be seen as a low-energy effective field theory \cite{Crisostomi:2017ugk}. 

\subsection{Comments for higher orders}
Let us discuss now operators of higher order. The quartic operators can be written as
{\allowdisplaybreaks  
\begin{align}
    L_1^{(2,2)} =& b_1 \left( C_{ij} C^{ij} - 2 C_i C^i + V_\star^2 \right) \left[ \bar R + K_{ij} K^{ij} + K^2 - 2 D_i a^i - 2 a_i a^i + 2  \nabla_n K \right], \\
    L_2^{(2,2)}=&  b_2 \left(  C_{ij} C^{ij} - 2 C_i C^i + V_\star^2  \right) \Big[ \bar R_{ij} D^i\varphi D^j\varphi + K K_{ij} D^i\varphi D^j\varphi + \nabla_n K_{ij} D^i\varphi D^j\varphi \nonumber \\ & - D_i a_j D^i\varphi D^j\varphi - a_i a_j D^i\varphi D^j\varphi - 2 A_\star D^i\varphi \left( D_j K^j_i - D_i K\right) - A_\star^2 K_{ij} K^{ij} + A_\star^2 D_i a^i  \nonumber \\  & + A_\star^2 a_i a^i - A_\star^2 \nabla_n K \Big], \\ 
    L_3^{(2,2)}=& b_3 \left( C_i^i - V_\star \right)^2 \left[  \bar R + K_{ij} K^{ij} + K^2 - 2 D_i a^i - 2 a_i a^i + 2 \nabla_n K \right], \\
    L_4^{(2,2)}=&  b_4 \left( C_i^i - V_\star \right)^2  \Big[ \bar R_{ij} D^i\varphi D^j\varphi + K K_{ij} D^i\varphi D^j\varphi +  \nabla_n K_{ij} D^i\varphi D^j\varphi  - D_i a_j D^i\varphi D^j\varphi  \nonumber \\ &  - a_i a_j D^i\varphi D^j\varphi - 2 A_\star D^i\varphi \left( D_j K^j_i - D_i K\right) - A_\star^2 K_{ij} K^{ij} \nonumber \\ & + A_\star^2 D_i a^i  + A_\star^2 a_i a^i - A_\star^2  \nabla_n K \Big], \\
    L_5^{(2,2)}=& b_5 \Big[  \bar R_{ijkl} C^{ik} C^{jl} + K_{ik} K_{jl} C^{ik} C^{jl} - K_{ij} K_{kl} C^{ik} C^{jl} + \left( 2 V_\star C^{ik} - 2 C^i C^k \right) \nonumber \\
    & \times \Big( - \nabla_n K_{ik} - K_{ij} K^j_k + D_i a_k + a_i a_k \Big) - 4 C^{ij} C^k \left( D_i K_{kj} - D_k K_{ij} \right) \Big], \\
    L_6^{(2,2)}=& b_6 \left( C_i^i - V_\star \right) \Big[ \bar R_{ij} C^{ij} + K K_{ij} C^{ij} + \nabla_n K_{ij} C^{ij} - D_i a_j C^{ij}  - a_i a_j C^{ij}  \nonumber \\ 
    &  -2 C^i \left( D_j K^j_i - D_i K\right) - V_\star K_{ij} K^{ij} + D_i a^i V_\star + a_i a^i V_\star - \nabla_n K V_\star 
    \Big].
\end{align}
}
The operator $L_3^{(2,2)}$ is again a specific subcase of the analysis in \cite{Zheng:2018cuc}. Further, from the previous discussion of the cubic operators we can note that the Hamiltonian analysis can be straightforwardly generalized to the combination of the two quartic operators $L_3^{(2,2)}$ and $L_4^{(2,2)}$. 

For the operators $L_1^{(2,2)}$ - $L_4^{(2,2)}$ we still have the same two scalar components of the extrinsic curvature which acquire a time derivative, namely $E$ and $F$. In the appendix \ref{section:Hamiltonian_analysis_quartic_order} we sketch the corresponding Hamiltonian analysis for these operators, arguing that by imposing the degeneracy condition \eqref{eq:Degeneracy_condition_quartic} the theory will in general have four dof which reduce to three if we take the homogeneous condition for the scalar field $\varphi\approx t$. 

In contrast, the last two operators $L_5^{(2,2)}$ and $L_6^{(2,2)}$ have more independent components of the extrinsic curvature with time derivatives. Therefore, we might expect more additional dof for generic scalar field configurations $\partial_i \varphi \not\approx 0$, which are removed by identifying the time flow with the scalar field. See, for instance, \cite{Crisostomi:2017ugk}, where the restriction to homogeneous field configurations removes two or more (Ostrogradski) ghost dof in modified Chern-Simons gravity models. However, the full Hamiltonian analysis is beyond the scope of this paper.

The same argumentation also applies to quintic or higher orders. Indeed, it is for instance straightforward to show that the analysis of the cubic operators carries on for the two quintic operators $L_1^{(3,2)}$ and $L_2^{(3,2)}$ as well. Note, that for the quintic order we have a new feature, due to the inclusion of operators which are quadratic in the curvature. Consequently, the terms with time derivatives of the extrinsic curvature are no longer just linear in the action. As a simple example, 
in the appendix \ref{section:Hamiltonian_analysis_quintic_order}
we will sketch the Hamiltonian analysis for the term \eqref{eq:Quadratic_operator_quintic} combined with the operator $L_1^{(3,2)}$, showing that there will be in general again four dof, reducing to three by imposing $\varphi\approx t$. 

\section{Non-homogeneous field configurations: Perturbations around Minkowski}
To understand better the nature of the additional scalar degrees of freedom in non-homogeneous field configurations, we consider as an example the action
\begin{align}
\label{eq:action_for_second_order_pert}
    S = \int \md^4x\,\sqrt{-g} \left( L_2^{(1,2)} - \lambda \left( g^{\mu\nu} \partial_\mu\varphi \partial_\nu +1 \right) \right).
\end{align}
In the following, we discuss the scalar perturbations around flat space-time $g_{\mu\nu}=\eta_{\mu\nu}$ for the homogeneous case $\varphi=t$ and the inhomogeneous case $\varphi= \sqrt{1+ a^i a_i}\; t + a_i x^i$. Note, that the inhomogeneous profile of the scalar field is the most general one consistent with the mimetic constraint in flat space-time and fulfills the full set of the EOM for a generic $a_i$ and a vanishing background value of the Lagrange parameter $\lambda_0 = 0$.

\subsection{Second-order action and gauge fixing}
Considering only scalar perturbations
\begin{align}
    \md s^2 = - ( 1 + 2 \Phi ) \md t^2 + 2 \partial_i B \md x^i \md t + \left( ( 1 - 2 \Psi ) \delta_{ij} + 2 \partial_i \partial_j E \right)\md x^i \md x^j 
\end{align}
the second-order of the action \eqref{eq:action_for_second_order_pert} without gauge fixing can be written as
\begin{align}
    S^{(2)}= \int \md^3x\,\md t\Big[& - 2 \Delta \Phi \Delta \delta\varphi + 4 \Delta \delta\varphi \Delta \Psi - 10 \dot\varphi \dot\Psi \Delta \Phi  + 2 \Delta\Phi \delta\ddot\varphi -10\Delta\Psi \delta\ddot\varphi  - 6 \Delta\delta\varphi \ddot\Psi - 6 \dot\varphi \dot\Phi \ddot\Psi \nonumber \\
    & + 6\delta\ddot\varphi \ddot\Psi + 2 \Delta\Psi \partial^i \varphi \partial_i \Phi - 6 \ddot\Psi \partial^i \varphi \partial_i\Phi - 2 \Delta \dot B \Delta \delta\varphi + 2 \Delta \ddot E \Delta\delta\varphi + 2 \Delta \ddot E \partial^i \varphi \partial_i \dot B  \nonumber \\
    & + 2 \Delta \Psi \partial^i \varphi \partial_i \dot B + 2 \Delta \ddot E \partial^i \varphi \partial_i \Phi + 2 \Delta \dot B \partial^i \varphi \Delta\partial_i E + 2 \Delta \Phi \partial^i \varphi \Delta\partial_i E - 4 \Delta \Psi \partial^i \varphi \Delta \partial_i E \nonumber \\
    & - 2 \dot\varphi \Delta \Phi \Delta B + 4 \dot\varphi \Delta\Psi \Delta B - 2 \dot\varphi \dot \Phi \Delta \dot B + 2 \dot \varphi \Delta \Phi \Delta \dot E - 4 \dot\varphi \Delta\Psi \Delta \dot E + 2\dot \varphi \dot \Phi \Delta \ddot E \nonumber \\
    & + 2 \Delta \dot B \delta\ddot \varphi - 2 \Delta \ddot E \delta\ddot \varphi - 6 \ddot\Psi \partial^i \varphi \partial_i \dot B + 8 \ddot \Psi \partial^i \varphi \Delta\partial_i E 
     - \delta\lambda \Big( 2 \dot\varphi^2 \Phi - 2 \dot \varphi \delta\dot\varphi \nonumber \\
    &  + 2 \partial^i \varphi \partial_i \delta\varphi + 2 \Psi (\partial_i\varphi)^2 - 2 \partial^i \varphi \partial^j\varphi \partial_i \partial_j E + 2 \dot \varphi \partial^i \varphi \partial_i B\Big) \Big].
\end{align}
Using the EOM for $\delta\lambda$ (i.e. using the mimetic constraint) we can express the gravitational potential $\Phi$ as 
\begin{align}
    \Phi = \frac{\delta\dot\varphi}{\dot\varphi} - \frac{(\partial_i \varphi)^2}{\dot\varphi^2} \Psi - \frac{\partial^i \varphi}{\dot\varphi^2} \partial_i \delta\varphi + \frac{\partial^i \varphi \partial^j\varphi}{\dot\varphi^2} \partial_i \partial_j E - \frac{\partial^i\varphi}{\dot\varphi} \partial_i B
\end{align}
Plugging it back into the action we obtain
\begin{align}
    S^{(2)} =\int \md^3 x\, \md t \Big[ & \frac{2}{\dot\varphi^2} \dot\Psi \left((\partial_i\varphi)^2 \Delta - 3 \partial^i \varphi \partial^j \varphi \partial_i \partial_j \right) \dot \zeta + 8 \dot\Psi \frac{1}{\dot\varphi} \partial^i \varphi \Delta\partial_i \zeta - \frac{2}{\dot\varphi^2} \partial^i\varphi \partial^j \varphi \partial_i\partial_j \zeta \Delta\Psi \nonumber \\
     & + \left( 4 + 2 \frac{(\partial_i \varphi)^2}{\dot\varphi^2} \right) \Delta\zeta \Delta\Psi \Big],
\end{align}
where we have introduced the Laplace operator $\Delta\equiv \delta^{ij} \partial_i \partial_j$ and
\begin{align}
    \zeta = \delta\dot\varphi + \dot\varphi B - \partial^k\varphi \partial_k E - \dot\varphi \dot E.
\end{align}
From the action above, we can observe that there is a gauge degree of freedom which can be used to set $E=0,\; B=0$ (Poisson Gauge) or $E=0,\;\delta\varphi=0$ (Unitary Gauge) without loss of generality. In the following we will use the Poisson Gauge.

\subsection{Homogeneous case $\varphi=t$}
In this case we obtain just
\begin{align}
    S^{(2)} = 4 \int \md^3x\, \md t\;  \Delta \Psi \Delta\delta\varphi
\end{align}
To obtain a propagating degree of freedom we consequently need to add additional terms. Further, to get a stable degree of freedom around the Minkowski background we have to break the shift symmetry of the scalar field to make the background explicitly time dependent. 
However, this would make the calculation in the non-homogeneous configuration very involved. Therefore, we will restrict ourselves to $L_2^{(1,2)}$, keeping in mind that this is a very specific example.

\subsection{Non-homogenous case $\partial_i \varphi\neq 0$}
For the general case the action can be written as
\begin{align}
     \delta S^{(2)} = \int \md^3x \md t \Big[& \frac{2}{\dot\varphi^2} \dot\Psi \left((\partial_i\varphi)^2 \Delta - 3 \partial^i \varphi \partial^j \varphi \partial_i \partial_j \right) \delta\dot\varphi + 8 \dot\Psi \frac{1}{\dot\varphi} \partial^i \varphi \Delta\partial_i \delta\varphi - \frac{2}{\dot\varphi^2} \partial^i\varphi \partial^j \varphi \partial_i\partial_j \delta\varphi \Delta\Psi \nonumber \\
     & + \left( 4 + 2 \frac{(\partial_i \varphi)^2}{\dot\varphi^2} \right) \Delta\delta\varphi \Delta\Psi
     \Big].
\end{align}
Now let us define the spatial derivative operator 
\begin{align}
    \tilde \Box \Psi \equiv \left(  (\partial_i \varphi)^2 \Delta - 3 \partial^i \varphi \partial^j \varphi \partial_i \partial_j \right) \Psi.
\end{align}
The specific form will explicitly depend on the form of our background field. In particular, note that the operator is in general neither negative nor positive definite. For simplicity, let us discuss the example $\varphi= \sqrt{1 + \alpha^2}\,t + \alpha x$:
\begin{align}
    \tilde \Box = \alpha^2 ( \partial_y^2 + \partial_z^2) - 2\alpha^2 \partial_x^2.
\end{align}
One can decouple the two degrees of freedom by performing a transformation of variable
\begin{align}
    u_+ = \Psi + \delta \varphi,\qquad u_- = \delta \varphi - \Psi
\end{align}
Then we can rewrite the second-order action as
\begin{align}
    \delta S = \int \md^4x \Big[ & \frac{1}{2\dot\varphi^2} \dot u_+ \tilde \Box \dot u_+ + \frac{2}{\dot\varphi} \dot u_+ \partial^i \varphi \Delta \partial_i u_+ + \left( 1 + \frac{(\partial_i \varphi)^2}{2 \dot\varphi^2} \right) \left( \Delta u_+ \right)^2 - \frac{1}{2 \dot\varphi^2} \partial^i\varphi \partial^j\varphi \partial_i \partial_j  u_+ \Delta u_+ \nonumber \\ & - \left( u_+ \Longleftrightarrow u_- \right)  \Big].
\end{align}
Therefore, at least one of the two dof is a ghost, which can also be seen from the Hamiltonian
\begin{align}
    \mathcal{H} =& \frac{\dot\varphi^2}{2} p_+ \tilde \Box^{-1} p_+ - 2 \dot \varphi \partial^i \varphi  \Delta \partial_i u_+ \tilde \Box^{-1} p_+ + 2 \partial^i \varphi \partial^j\varphi \Delta \partial_i u_+ \tilde \Box^{-1} \Delta \partial_j u_+ -  \left( 1 + \frac{(\partial_i \varphi)^2}{2 \dot\varphi^2} \right) \left( \Delta u_+ \right)^2 \nonumber \\ & 
    + \frac{1}{2 \dot\varphi^2} \partial^i\varphi \partial^j\varphi \partial_i \partial_j  u_+ \Delta u_+ - \left( u_+,\; p_+ \Longleftrightarrow u_-,\; p_- \right) .
\end{align}
Note that, since the operator $\tilde \Box$ is in general not sign-definite, even for both of the dof the kinetic energy can be negative. The EOM are
\begin{align}
    & \frac{1}{\dot\varphi^2} \tilde \Box \ddot u_{+/-} + 4 \frac{\partial^i\varphi}{\dot\varphi} \Delta \partial_i \dot u_{+/-} + \frac{1}{\dot\varphi^2} \partial^i \varphi \partial^j \varphi \Delta\partial_i \partial_j u_{+/-} - \left(2 + 1 \frac{(\partial_i\varphi)^2}{\dot\varphi^2} \right) \Delta \Delta u_{+/-}=0.
\end{align}
The dispersion relation depends on the form of the background field, due to $\partial_i\varphi$. Considering the example 
$\varphi=\sqrt{1+\alpha^2 }\, t + \alpha x$ the dispersion relation can be written as
\begin{align}
   \alpha^2 \omega^2 \left( _\bot k^2 - 2 k_x^2 \right) - 4 \omega \alpha \sqrt{1+\alpha^2} \left( _\bot k^2 + k_x^2 \right) k_x + \alpha^2 \left( _\bot k^2 + k_x^2 \right) k_x^2 - ( 3\alpha^2 + 2) ( _\bot k^2 + k_x^2 ) ^2 = 0,
\end{align}
where $_\bot k^2 = k_y^2 + k_z^2$. We can observe that there are no gradient instabilities, namely $\omega \in \mathbf{R}$. Therefore, the modes are by themselves stable. However, the coupling between $u_+$ and $u_-$ at higher order may introduce instabilities.

Summarizing, we can observe that at the level of linear perturbations the additional scalar dof can only be seen if the background scalar field is time and space-dependent, consistently with similar results in \cite{DeFelice:2018mkq,Iyonaga:2018vnu,Blas:2009yd}.

\section{Summary and Conclusion}

In this paper we have analyzed the direct coupling of higher derivative operators of the scalar field to the curvature in mimetic gravity theories which were proposed to solve the instability problems for linear perturbations around the FLRW background. Imposing homogeneous field configurations $\varphi=t$ at the action level we have derived degeneracy conditions between the different operators to obtain a gravity theory with three dof up to the quintic order. 

In contrast, we have shown for the cubic operators, by performing the full Hamiltonian analysis for generic scalar field configurations, that these mimetic gravity theories have in general four instead of three dof. This discrepancy can be explained by the fact that the Dirac matrix is singular in the homogeneous point $ \partial_i \varphi\approx 0$, which is in agreement with similar results in \cite{Zheng:2018cuc}. Therefore, imposing the homogeneous condition via the constraint $C_t = \varphi -t \approx 0$ does not just lead to the standard gauge fixing conditions but also to two further second-class constraints which reduce the number of dof by one. Therefore, one has to distinguish two different branches of solutions with $\partial_i \varphi \not \approx 0$ and $\partial_i \varphi\approx 0$.
While the homogeneous point is singular, the theory in this singular point is itself well defined and can be seen as a Lorentz violating theory or a low-energy effective field theory (similar discussions \cite{Gomes:2017tzd,Crisostomi:2017ugk}).
The results can be extended to higher orders, but, due to the presence of further components of the extrinsic curvature with time derivatives, for some of these operators there might be in general even more than four dof, which reduce to three by imposing the singular homogeneous condition. This analysis is postponed to a future project.

Finally, we have discussed the linear perturbations around a Minkowski background for a homogeneous, $\varphi=t$ and non-homogeneous $\varphi=\sqrt{1+\alpha^2}\,t+\alpha x$ background scalar field, for one particular choice of the higher-derivative operator to understand better the presence of the additional scalar degree of freedom. We have observed that, for the non-homogeneous background scalar field, there are two propagating scalar degrees of freedom present, as it is expected from the Hamiltonian analysis for generic scalar fields. For this specific example the two scalar dof decouple from each other and at least one of them has a ghost-like behaviour (negative kinetic energy). By analyzing the dispersion relation we have shown that there are no gradient instabilities and the modes themselves are stable. However, instabilities may appear at higher orders, due to the coupling of the two scalar degrees of freedom. 

Summarizing, the restriction of homogeneous field configurations operators with direct coupling of higher derivatives to the curvature can provide new well-defined (no Ostrogradski ghost) explicitly Lorentz violating mimetic gravity theories. In the future, we plan to study the phenomenological properties of these models and their possible cosmological applications.

\acknowledgments

We would like to thank G. Tasinato, P. Karmakar and especially D. Sorokin for useful discussions and comments on a preliminary version of this paper. This work is partially supported by 
ASI Grant No. 2016-24-H.0. Part of the computations are done using Mathematica\footnote{https://www.wolfram.com/mathematica/}, with the algebra package xAct\footnote{http://www.xact.es/}.

%

\appendix

\section{Hamiltonian analysis of $L_1^{(2,2)}$ in the homogeneous field configuration}
\label{sec:Appendix_Hamiltonian_analysis_Unitary_Gauge_L1}
The action for the quartic operator $L_1^{(2,2)}$ can be rewritten as
\begin{align}
    S &=\int d^4x\,\sqrt{-g} \left[  S^2 - E^2 S + S \bar R - 2 \nabla_n S E + \Lambda \left( S - E_{ij} E^{ij} \right)\right]
\end{align}
The canonical conjugate momenta are given by
\begin{align}
    \pi_{ij} = \frac{1}{2} \frac{\delta L}{\delta E^{ij}} = - \sqrt{h} S E h_{ij} - \sqrt{h} h_{ij} \nabla_n S - \sqrt{h} \Lambda E_{ij}, \quad p_S = - 2 E \sqrt{h} 
\end{align}
The others, $\pi_i$ and $p_\Lambda$, are trivial primary constraints. Inversion yields
\begin{align}
    \nabla_n S = - \frac{\pi }{3 \sqrt{h}} + \frac{p_S}{2 \sqrt{h}} \left( S + \frac{1}{3} \Lambda  \right), \quad E_{ij} = - \frac{1}{\sqrt{h} \Lambda} \left( \pi_{ij} - \frac{1}{3} \pi h_{ij} \right)- \frac{p_S}{6 \sqrt{h}} h_{ij}
\end{align}
From this we get the extended Hamiltonian 
\begin{align}
    H_T = \int d^3x\, \left( \mathcal{H} + N^i \mathcal{H}_i + u^i \pi_i + u_\Lambda p_\Lambda  \right),
\end{align}
with
\begin{align}
    \mathcal{H} &= - \frac{1}{\Lambda \sqrt{h}} \left( \pi_{ij} \pi^{ij} - \frac{1}{3} \pi^2 \right) - \frac{1}{3} \frac{\pi p_S}{\sqrt{h}} + \frac{1}{4} \frac{p_S^2}{\sqrt{h}} \left( S + \frac{1}{3} \Lambda \right) - \sqrt{h} S^2 - \sqrt{h} S \bar R - \sqrt{h} S \Lambda, \\
    \mathcal{H}_i &= - 2 D_j \pi^j_i + p_S \partial_i S + p_\Lambda \partial_i \Lambda
\end{align}
As usual, the conservation of $\pi_i$ yields the constraint $\mathcal{H}_i$. Together, $\pi^i$ and $\mathcal{H}_i$ form the six first-class constraints, corresponding to the invariance under time-independent spatial transformations.

The conservation of $p_\Lambda$ yields
\begin{align}
    \dot p_\Lambda = \{p_\Lambda, H_T\} \approx - \frac{1}{12} \frac{p_S^2}{\sqrt{h}} - \frac{1}{\sqrt{h}\Lambda^2} \left( \pi_{ij} \pi^{ij} - \frac{1}{3} \pi^2 \right) + \sqrt{h} S \equiv  C_\Lambda
\end{align}
The conservation of $C_\Lambda$ fixes the Lagrange parameter $u_\Lambda$ 
\begin{align}
    \dot C_\Lambda \approx - \frac{1}{\sqrt{h} \Lambda^3} \left( \pi_{ij} \pi^{ij} - \frac{1}{3} \pi^2 \right) u_\Lambda + \{ C_\Lambda , \int \md^3x \mathcal{H} \}.
\end{align} 
Therefore, the chain of constraints ends here. Finally, there are six 
first-class constraints, corresponding to the spatial transformation, $\pi_i$ and $\mathcal{H}_i$, and two second-class constraints, $p_\Lambda$ and $C_\Lambda$, which results in four degrees of freedom.

\section{Degeneracy in the homogeneous field configuration}
\label{sec:Degeneracy_1_3_model}
To analyze the behaviour of the mimetic gravity model which are degenerate in the homogeneous field configuration $\varphi=t$, it is instructive to discuss for simplicity the following model 
\begin{align}
    S=& \int \md^4x\,\sqrt{-g} \left[ \frac{1}{2} R + \frac{1}{3} \left( \Box \varphi\right)^2 - \lambda \left( \partial_i \varphi \partial^i \varphi +1 \right) \right] \nonumber \\
    =&  \int \md^4x\,\sqrt{-g}\,\Big[  \frac{1}{2} R +   \frac{1}{3} \chi^2 + \epsilon \chi+ g^{\mu\nu} \partial_\mu \epsilon \partial_\nu \varphi -\lambda \left(g^{\mu\nu}\partial_\mu\varphi\partial_\nu\varphi + 1\right)\Big].
\end{align}
As discussed in \cite{Zheng:2018cuc,Ganz:2018mqi}, there are in general three degrees of freedom. However, restricting to the homogeneous field configuration $\varphi\approx t$ the number of degrees of freedom reduces to two. The reason for this reduction is the same as in the models discussed in this paper. 

Therefore, it is instructive to analyze the origin of this reduction in more detail. The Hamiltonian constraint derived in \cite{Ganz:2018mqi} has the following form
\begin{align}
    \mathcal{H} =& \mathcal{H}_{\mathrm{gr}} - \lambda \frac{p_\epsilon^2}{\sqrt{h}} -  \frac{p_\epsilon p_\varphi}{\sqrt{h}} +  \lambda \left(h^{ij}\partial_i\varphi \partial_j\varphi +1\right) -  \frac{1}{3} \sqrt{h} \chi^2 -  \sqrt{h} \epsilon \chi - \sqrt{h} h^{ij}\partial_i\epsilon \partial_j\varphi,
\end{align}
where $\mathcal{H}_{\mathrm{gr}}$ is the standard Hamiltonian constraint of GR.
There are six second-class constraints, which can be expressed as $p_\chi \approx 0$, $p_\lambda \approx 0$ and
{\allowdisplaybreaks
\begin{align}
    C_\chi &= \frac{2}{3} \chi +  \epsilon, \\
 C_\lambda^{(1)} &= \left( -  \sqrt{h} \left(h^{ij}\partial_i\varphi \partial_j\varphi +1\right) + \frac{ p_\epsilon^2}{\sqrt{h}}\right) , \\
C_{\lambda}^{(2)} &\approx \left(- 3 p_\epsilon \epsilon - 2 \frac{p_{\epsilon}}{\sqrt{h}}\partial_i \left(\sqrt{h} h^{ij} \partial_j\varphi \right) +2 \sqrt{h} h^{ij}\partial_i\varphi \partial_j \left(\frac{p_\epsilon}{\sqrt{h}}\right) + 4 \pi^{ij} \partial_i\varphi \partial_j\varphi + 2 \pi \right) , \\
C_\lambda^{(3)} \approx &  - 2 \lambda \sqrt{h} h^{ij} \partial_i\varphi \partial_j \varphi  - 2 D_i \varphi D^i\varphi \frac{p_\epsilon p_\varphi}{\sqrt{h}} + \frac{1}{2} \sqrt{h} \bar{R}  \nonumber \\ & + 2 \epsilon \chi \sqrt{h}   D_i\varphi D^i \varphi + 4 \frac{p_\epsilon}{\sqrt{h}} D_i D^i p_\epsilon - 4 \sqrt{h} \chi D_i D^i \varphi + 6 \frac{\pi }{\sqrt{h}} D_i \varphi D^i p_\epsilon  \nonumber \\ & - 3 \sqrt{h} D_i \varphi D^i \epsilon + \sqrt{h} D_i\varphi D^i \epsilon   +\sqrt{h}\bar{R} D_i\varphi D^i\varphi  + 4 \frac{\pi^{kl}\pi_{kl}}{\sqrt{h}} D_i \varphi D^i \varphi   \nonumber \\ & + \sqrt{h} \chi^2 ( 1 + \frac{2}{3} D_i\varphi D^i \varphi) - \frac{\pi^2}{2\sqrt{h}} ( 6 + 4 D_i \varphi D^i \varphi) + 8 \frac{p_\epsilon}{\sqrt{h}} D^i \varphi D_j \pi^{j}_i + 8 \frac{p_\epsilon \pi^{ij}}{\sqrt{h}} D_i D_j \varphi \nonumber \\ &  + 2\sqrt{h} \left(D_i D^i \varphi \right)^2 - 4 \sqrt{h} D^i \varphi D_j D^j D_i\varphi - 16 \frac{\pi_i^k\pi_{jk}}{\sqrt{h}} D^i \varphi D^j\varphi -2 \sqrt{h} D_i\varphi D^i \epsilon D_j \varphi D^j\varphi \nonumber \\ & - 2 \frac{p_\epsilon \pi }{\sqrt{h}} D_i D^i \varphi + 6 \frac{\pi^{ij}\pi_{ij}}{\sqrt{h}} - 16 \frac{\pi^{ij}}{\sqrt{h}} D_i p_\epsilon D_j\varphi + 8 \frac{\pi \pi^{ij}}{\sqrt{h}} D_i\varphi D_j\varphi - 2 \sqrt{h} D_i D_j\varphi D^i D^j \varphi.
\end{align}
}
The constraint $C_\lambda^{(3)}$ depens explicitly on $\lambda$ and its conservation fixes the Lagrange parameter $u_\lambda$ associated to the primary constraint $p_\lambda$. 
The Dirac matrix $\Omega_{IJ}=\{C_I,C_J\}$, where $C_I$ are the six second class constraints $p_\lambda,\;p_\chi,\;C_\chi,$ $C_\lambda^{(1)},\;C_\lambda^{(2)}$ and $C_\lambda^{(3)}$, can be expressed as
\begin{align}
\Omega_{IJ} =
\begin{pmatrix}
0 & 0 & 0 & 0 & 0 & A \\
0 & 0 & B & 0 & 0 & C \\
0 & -B & 0 & D & E & F \\
0 & 0 & -D & 0 & M & H \\
0 & 0 & -E & -M & I & J \\
-A & - C & -F & -H & -J & -K 
\end{pmatrix}.
\end{align}
Consequently, the determinant is given by
\begin{align}
\label{eq:Dirac_bracket}
\det \Omega_{IJ} = A^2 B^2 M^2.
\end{align}
The Dirac matrix is only well defined and invertible if the determinant is weakly non-vanishing, which means that
\begin{align}
A &= \{p_\lambda(x) , C_\lambda^{(3)}(y)\} = 2 \sqrt{h} h^{ij} \partial_i \varphi \partial_j \varphi \delta^{(3)}(x-y) , \\
B &=\{p_\chi(x),C_\chi(y) \} = - \frac{2}{3} \delta^{(3)}(x-y) , \\
M &= \{C_\lambda^{(1)}(x),C_\lambda^{(2)}(y)\} \approx-2 \sqrt{h} h^{ij}\partial_i \varphi \partial_j\varphi \delta^{(3)}(x-y),
\end{align}
has to be weakly non-vanishing. This shows directly that for any solution with $D_i \varphi \approx 0$ the determinant vanishes weakly. On the other hand, we could note that for $D_i \varphi \approx 0$ the second-class constraint $C_\lambda^{(3)}$ does not depend on $\lambda$ anymore and consequently its conservation does not fix the Lagrange paramter $u_\lambda$ and its evolution can evolve to non-vanishing values. 
Consequently, we have to distinguish two scenarios with $D_i \varphi \approx 0$ and $D_i \varphi \not\approx 0$. 

Let us focus now on the first case. For this reason let us introduce the constraint
\begin{align}
\label{eq:C_t}
    C_t = \varphi - t \approx 0.
\end{align}
Its time evolution yields another constraint, namely
\begin{align}
\label{eq:C_t_2}
    C_t^{(2)}= - 1 - N \frac{p_\epsilon}{\sqrt{h}} \approx 0.
\end{align}
These two constraints do not commute with $\pi_N$ and $\mathcal{H}$ and consequently fix the gauge freedom corresponding to the time-diffeomorphism invariance. From this point of view the constraints $C_t$ and $C_t^{(2)}$ could be understood as a gauge-fixing condition $\varphi=t$, which is normally called unitary gauge.

While in standard gravity models this is indeed the case, we have already seen before that for this model the choice of $\varphi\approx t$ is singular, as far as the Dirac bracket \eqref{eq:Dirac_bracket} is concerned. This requires further analysis. Indeed, checking the chain of constraints we note that $C_\lambda^{(3)}$ is now given by
\begin{align}
    C_\lambda^{(3)} \approx &    \frac{1}{2} \sqrt{h} \bar{R}     - 2 \frac{\pi^2}{\sqrt{h}} + 6 \frac{\pi^{ij}\pi_{ij}}{\sqrt{h}} \approx 0.
\end{align}
As outlined before, $C_\lambda^{(3)}$ does not depend on $\lambda$ anymore and the chain of constraints does not stop. Instead the conservation of $C_\lambda^{(3)}$ leads to two further second-class constraints which are given by
\begin{align}
    C_\lambda^{(4)} \approx & - 8 \bar R_{ij} \pi^{ij} + 2 \pi \bar R \approx 0, \\
    C_\lambda^{(5)} \approx & 2 \sqrt{h} \bar R \lambda - p_\varphi \bar R + 4 \sqrt{h} \bar R_{ij} \bar R^{ij} + 16 \bar R \frac{\pi^{ij} \pi_{ij}}{\sqrt{h}} + 32 \bar R^{ij} \frac{\pi_i^k \pi_{jk}}{\sqrt{h}} - \frac{20}{3} \bar R \frac{\pi^2}{\sqrt{h}} + 16 \frac{\pi^{ij}}{\sqrt{h}} D_k D^k \pi_{ij} \nonumber \\ 
    & - 8 \frac{\pi}{\sqrt{h}} D_k D^k \pi - 32 \frac{\pi^{ij}}{\sqrt{h}} D_k D_j \pi_i^k + 8 \frac{\pi}{\sqrt{h}} D_j D_k \pi^{kj} + 8 \frac{\pi^{ij}}{\sqrt{h}} D_i D_j \pi.
\end{align}
Since $C_\lambda^{(5)}$ explicitly depends on $\lambda$ its time conservation fixes the Lagrange parameter $u_\lambda$ and the chain of constraints stops here.

\section{Hamiltonian analysis of quartic operators}
\label{section:Hamiltonian_analysis_quartic_order}
Considering the operators $L_1^{(2,2)}$ - $L_4^{(2,2)}$ and using the degeneracy condition \eqref{eq:Degeneracy_condition_quartic} the Lagrangian can be formally written as
\begin{align}
\label{eq:Lagrangia_quartic_order}
    L_{\mathrm{tot}}= L_{\mathrm{kin}} + L_0 - \lambda \left(-A_\star^2 + D_i \varphi D^i\varphi +1 \right),
\end{align}
with
\begin{align}
    L_{\mathrm{kin}}=& 2 b_1 \left( C_{kl} C^{kl} - C_k C^k + V_\star^2 \right) \left[ \left( 1- A_\star^2 + D_i \varphi D^i\varphi \right) V_E+ \left( 1 - A_\star^2\right) V_F \right] \nonumber \\
    & + \left( C_k^k - V_\star \right)^2 \left[ 2 b_3 \left( V_E + V_F \right) + b_4 \left(D_k \varphi D^k\varphi - A_\star^2 \right) V_E - b_4 A_\star^2 V_F \right].
\end{align}
Note, that in the first term the pre-factor of $V_E$ is proportional to the mimetic constraint. This will be later crucial for the number of dof.

Using the same notation and auxiliary variables as in subsection \ref{subsec:Hamiltonian_analysis_cubic_order} the non-trivial primary constraints can be written as
\begin{align}
    \bar P_E =& 2 b_1 \left( C_{kl} C^{kl} - C_k C^k + V_\star^2 \right)  \left( 1- A_\star^2 + D_i \varphi D^i\varphi \right)  + \left( C_k^k - V_\star \right)^2 \nonumber \\ &\times  \left[ 2 b_3 + b_4 \left(D_k \varphi D^k\varphi - A_\star^2 \right) \right] - P_E \approx 0, \\
    \bar P_F =& 2 b_1 \left( C_{kl} C^{kl} - C_k C^k + V_\star^2 \right) \left( 1 - A_\star^2\right) + \left( C_k^k - V_\star \right)^2 \left[ 2 b_3 - b_4 A_\star^2 \right] - P_F \approx 0,
\end{align}
while the other non-trivial primary constraints are given by \eqref{eq:Primary_constraint_p_f} - \eqref{eq:Primary_constraint_Pi}. The Hamiltonian constraint is given by
\begin{align}
\label{eq:Hamiltonian_constraint_quartic_order}
   \mathcal{H} =& - L_0 + \lambda ( - A_\star^2 + D_k\varphi D^k\varphi + 1) + p_\varphi A_\star + 2 _\bot{\hat{\tilde B}^{ij}} _\bot{\hat{\tilde \pi}_{ij}} + 2 \hat B^{ij} \hat \pi_{ij} + 2 \pi_E E + 2 \pi_F F \nonumber \\
    & + p_\star V_\star - \sqrt{h} D_i\left( \frac{p_\star}{\sqrt{h}} D^i\varphi \right).
\end{align}
The time conservation of $p_\lambda$ leads to the mimetic constraint $C_\lambda$ \eqref{eq:Mimetic_constraint}. On the other hand, from the time conservation of $\bar P_E$, $\bar P_F$, $P_{V_\star}$, $(_\bot\hat{\tilde P}_B)_{ij}$ and $(\hat P_B)_{ij}$ we obtain
\begin{align}
    \frac{\md }{\md t} P_{V_\star} \approx & \{ P_{V_\star}, \int \md^3x\, \bar P_E \} u_E + \{ P_{V_\star}, \int \md^3x\, \bar P_F \} u_F + \{ P_{V_\star}, \int \md^3x\, N \mathcal{H} \}, \\
    \frac{\md }{\md t} \bar P_E \approx & \{ \bar P_E , \int\md^3x\, P_{V_\star} \} u_{V_\star} + \{ \bar P_E, \int \md^3x\, \bar P_F \} u_F + \{ \bar P_E , \int \md^3x\, N\mathcal{H} \}, \\
     \frac{\md }{\md t} \bar P_F \approx & \{ \bar P_F , \int\md^3x\, P_{V_\star} \} u_{V_\star} + \{ \bar P_F, \int \md^3x\, \bar P_E \} u_E + \{ \bar P_F , \int \md^3x (_\bot\hat{\tilde P}_B)_{ij} \} (_\bot\hat{\tilde u}_B)^{ij}   \nonumber \\
     & + \{ \bar P_F , \int \md^3x\, (\hat P_B)_{ij} \} (\hat u_B)^{ij}   + \{ \bar P_E , \int \md^3x\, N\mathcal{H} \}, \label{eq:Conservation_P_F} \\
     \frac{\md}{\md t} (_\bot\hat{\tilde P}_B)_{ij} \approx & \{ (_\bot\hat{\tilde P}_B)_{ij}, \int \md^3x\, \bar P_F \} u_F + \{ (_\bot\hat{\tilde P}_B)_{ij}, \int\md^3x\, N \mathcal{H} \}, \\
     \frac{\md}{\md t} (\hat P_B)_{ij} \approx & \{ (\hat P_B)_{ij}, \int \md^3x\, \bar P_F \} u_F + \{ (\hat P_B)_{ij}, \int\md^3x\,N\mathcal{H} \},
\end{align}
where we have used the fact that all the commutators between the primary constraints are proportional to the delta-function, $\delta(x-y)$, (no spatial derivatives) so that we can pull out the Lagrange parameters. We can use the first two equations to fix $u_E$ and $u_{V_\star}$. Plugging it into equation \eqref{eq:Conservation_P_F} we obtain a relation between $(\hat u_B)_{ij}$ and $(_\bot\hat{\tilde u}_B)_{ij}$ so that it also does not lead to a secondary constraint. 

On the other hand, from the last two equations we can fix $u_F$ by multiplying for instance the time conservation of $(\hat P_B)_{ij}$ with $\hat B^{ij}$. Plugging back the solution for $u_F$ we obtain three independent secondary constraints, which we denote as $(\hat C_B)_{ij}$ and $(_\bot\hat{\tilde C}_B)_{ij}$. Their time evolution will fix the remaining components of $(\hat u_B)_{ij}$ and $(_\bot\hat{\tilde u}_B)_{ij}$.

The time conservation of the mimetic constraint leads to a tertiary constraint $C_\lambda^{(2)}$ which is given by \eqref{eq:C_lambda_2_cubic}. Its time evolution yields 
\begin{align}
    \dot C_\lambda^{(2)} \approx - 2 A_\star u_{V_\star} + 2 D_k\varphi D^k\varphi u_E + \{ C_\lambda^{(2)}, \int \md^3x\, N \mathcal{H} \}.
\end{align}
Using the expressions for $u_E$ and $u_{V_\star}$ we obtain a new constraint which, however, does not depend on $\lambda$ or $p_\varphi$. Further, the dependence on $p_\star$ and $\pi_E$ is given by the specific combination
\begin{align}
    C_\lambda^{(3)} \approx \frac{1}{2(2b_3 - b_4)} \frac{1}{C_k^k-V_\star} \left( 2 A_\star p_\star - 4 \pi_E D^k\varphi D_k\varphi \right) + ...\,, 
\end{align}
which commutes with the mimetic constraint $C_\lambda$. Consequently, the new constraint $C_\lambda^{(4)}$ due to the time conservation of $C_\lambda^{(3)}$ also does not depend on $\lambda$. The time conservation of $C_\lambda^{(4)}$ finally leads to $C_\lambda^{(5)}$. 

Note, that in general the operator $L_3^{(2,2)}$ is independent from the others. For $b_1=0$ and $b_4=0$ we have to recover the result from \cite{Zheng:2018cuc} that the theory has four dof. Together with $C_\lambda^{(5)}$ there are already 32 second class and eight first class constraints which would result in four dof. Consequently, we can deduce that in general $C_\lambda^{(5)}$ has to depend on $\lambda$ so that the chain of constraints stops.

When the two constraints $C_t$ and $C_t^{(2)}$  eqs. \eqref{eq:C_t} and \eqref{eq:C_t_2} are imposed, $\bar P_F$ commutes with $\bar P_E$, $(\hat P_B)_{ij}$ and $(_\bot\hat{\tilde P}_B)_{ij}$. Therefore, the time conservation of these operators leads to two further constraints, which have the same fundamental structure as in subsection \ref{subsec:Hamiltonian_analysis_cubic_order}. At the end there will be two further second-class constraints, reducing the number of dof by one, consistently with our previous discussions.

\section{Hamiltonian analysis of quintic operators}
\label{section:Hamiltonian_analysis_quintic_order}
 As an example, let us consider the following action 
\begin{align}
    S =& \int \md^4x\,\sqrt{-g} \Big[  d\, \Box\varphi \left( \frac{1}{2} R + R_{\mu\nu} \varphi^\mu \varphi^\nu \right)^2 + c\, ( \Box\varphi)^3 R - \lambda ( X + 1 ) \Big] \nonumber \\
    = & \int \md^4x\,\sqrt{-g} \Big[ d\, \Box \varphi \chi^2 + c\, (\Box\varphi)^3 R + \Omega \left( \chi - \frac{1}{2}R - R_{\mu\nu} \varphi^\mu \varphi^\nu \right) - \lambda ( X +1 )   \Big].
\end{align}
For simplicity, we assume that $d$ and $c$ are constant. The analysis is  very involved and we will just sketch the most important steps. We can split the Lagrangian formally as in \eqref{eq:Lagrangia_quartic_order}
\begin{align}
    L_{\mathrm{tot}}= L_{\mathrm{kin}} + L_0 - \lambda \left(-A_\star^2 + D_i \varphi D^i\varphi +1 \right),
\end{align}
with 
\begin{align}
    L_{\mathrm{kin}}=& 2 c \left( C_k^k - V_\star \right)^3 (V_E + V_F) - \Omega \Big[ (1- A_\star^2+ D_i\varphi D^i\varphi ) V_E + ( 1- A_\star^2) V_F \Big]. 
\end{align}
Using the same procedure as in section \ref{subsec:Hamiltonian_analysis_cubic_order} the non-trivial primary constraints can be written as
\begin{align}
    \bar P_F =& 2 c \left(C_k^k -V_\star \right)^3 - ( 1- A_\star^2) \Omega - P_F, \\
    \bar P_E =& 2 c \left(C_k^k -V_\star \right)^3 - ( 1+ D_k\varphi D^k\varphi- A_\star^2) \Omega - P_E,
\end{align}
while the other non-trivial primary constraints are given by \eqref{eq:Primary_constraint_p_f} - \eqref{eq:Primary_constraint_Pi}. The Hamiltonian constraint can be formally expressed as 
\begin{align}
    \mathcal{H} =& - L_0 + \lambda ( - A_\star^2 + D_k\varphi D^k\varphi + 1) + p_\varphi A_\star + 2 _\bot{\hat{\tilde B}^{ij}} _\bot{\hat{\tilde \pi}_{ij}} + 2 \hat B^{ij} \hat \pi_{ij} + 2 \pi_E E + 2 \pi_F F \nonumber \\
    & + p_\star V_\star - \sqrt{h} D_i\left( \frac{p_\star}{\sqrt{h}} D^i\varphi \right).
\end{align}
The time conservation of the primary constraints leads to the mimetic constraint $C_\lambda$ \eqref{eq:Mimetic_constraint} and
\begin{align}
   \frac{1}{N} \dot P_\chi \approx & - 2 d \left( C_k^k -V_\star \right) \chi - \Omega \equiv C_\chi \approx 0, \\
   \frac{1}{N}\frac{\md}{\md t} (_\bot\hat{\tilde P}_B)_{ij} \approx & 2 \left[ c \left( C_k^k - V_\star \right)^3  - \Omega \left( \frac{1}{2} - A_\star^2\right) \right]  { _\bot \hat{\tilde B}_{ij}}   - 2 \Omega  A_\star \left( P^k_i P^l_j - \frac{1}{3} h_{ij} P^{kl} \right) D_k D_l \varphi \nonumber \\ & - 2  _\bot\hat{\tilde \pi}_{ij} 
      \equiv  (_\bot \hat{\tilde C}_B)_{ij} \approx 0, \\
     \frac{1}{N} \frac{\md}{\md t}( { \hat P}_B)_{ij} \approx & 2 \Big[ c \left( C_k^k - V_\star \right)^3 - \Omega \left( \frac{1}{2} - A_\star^2\right) \Big] \hat B_{ij} + \Omega  P^k_{(j}  D_{i)} \varphi  D_{k} A_\star  - 2 \Omega D_{(i} \varphi \hat B^k_{j)} D_k \varphi  \nonumber \\ 
     & - 2  \sqrt{h}   \frac{ P^k_{(j} D_{i)} \varphi D^l \varphi}{D^m \varphi D_m \varphi } D_{(k} \left( \frac{1}{\sqrt{h}} \Omega A_\star D_{l)} \varphi \right)- 2 \hat \pi_{ij} \equiv (\hat C_B)_{ij} \approx 0.
\end{align}
Further, the time conservation of $P_\Omega$, $\bar P_F$, $\bar P_E$ and $P_{V_\star}$ fixes the associated Lagrange parameter $u_\Lambda$, $u_F$, $u_E$ and $u_{V_\star}$.

The time conservation of the secondary constraints $C_\chi$, $(_\bot\hat{\tilde C}_B)_{ij}$, $(\hat C_B)_{ij}$ fixes $u_\chi$, $(_\bot\hat{\tilde u}_B)_{ij}$, $(\hat u_B)_{ij}$. On the other hand, the time conservation of the mimetic constraint leads to a tertiary constraint $C_\lambda^{(2)}$ \eqref{eq:C_lambda_2_cubic}. Its time evolution yields a new constraint
\begin{align}
  N  C_\lambda^{(3)} \approx & 2 \sqrt{h} A_\star \{ \bar P_E, \int \md^3x\, N \mathcal{H} \} + 6  \sqrt{h}\,c \left( C_k^k - V_\star \right)^2 \{ P_\Omega , \int \md^3x\, N \mathcal{H} \} \nonumber \\ 
  &+ 2 \sqrt{h} D_k \varphi D^k\varphi \{ P_{V_\star}, \int \md^3x\, N \mathcal{H} \} - 3 \sqrt{h}\, c \left( C_k^k - V_\star \right)^2 \{ C_\lambda^{(2)} ,\int \md^3x\, N \mathcal{H} \}.
\end{align}
Contrary to the cubic analysis and similar to the analysis in appendix \ref{section:Hamiltonian_analysis_quartic_order}, it does not depend on $\lambda$ and $p_\varphi$. Further, it only depends on a special combination of $p_\star$ and $\pi_E$ 
\begin{align}
    C_\lambda^{(3)} \approx 2 \sqrt{h} \left( 2 A_\star \pi_E - p_\star D_i \varphi D^i\varphi \right) + ...\,,
\end{align}
which commutes with the mimetic constraint $C_\lambda$. Consequently, the new constraint $C_\lambda^{(4)}$ obtained by requiring the time conservation of $C_\lambda^{(3)}$ also does not depend on $\lambda$. Its time evolution finally leads to $C_\lambda^{(5)}$. Note that the two operators $L^{(1,4)}$ and $L_1^{(3,2)}$ are in general independent and for $d=0$ and $c\neq 0$ we recover the result from \cite{Zheng:2018cuc}. Using the same argumentation as in appendix \ref{section:Hamiltonian_analysis_quartic_order} we can deduce that in general $C_\lambda^{(5)}$ has to depend on $\lambda$ so that the chain of constraints stops here and we obtain four degrees of freedom. 

Imposing the two constraints $C_t$ and $C_t^{(2)}$ eqs. \eqref{eq:C_t} and \eqref{eq:C_t_2} we obtain the same results as in the other cases. The chain of constraints breaks down, since the conservation of $P_\Omega$, $\bar P_E$, $\bar P_F$ and $P_{V_\star}$ is not sufficient anymore to fix the corresponding Lagrange parameters and consequently we have to impose further constraints. At the end there will be two further second-class constraints reducing the number of the dof to three.

\bibliography{bibliography}
\bibliographystyle{JHEP}

\end{document}